\documentclass[aps,pre,reprint,showpacs,superscriptaddress,groupaddress,showkeys,floatfix,byrevtex3]{revtex4-1}
\usepackage{graphicx}
\usepackage{dcolumn}
\usepackage{bm}
\usepackage{srctex}
\usepackage{stackrel}
\usepackage{amsmath}
\usepackage{amsfonts}
\usepackage{amssymb}
\usepackage[dvipsnames]{xcolor}
\usepackage{subfigure}
\usepackage{cancel}
\usepackage{soul}

\begin{document}
 \title{Fractional and scaled Brownian motion on the sphere: The effects of long-time correlations on navigation strategies}

\author{Adriano \surname{Vald\'es G\'omez}}
\email[]{adriano@ciencias.unam.mx}
\thanks{corresponding author}
\affiliation{Facultad de Ciencias, Universidad Nacional Aut\'onoma de M\'exico, Alcaldía Coyoacán, C.P. 04510, Ciudad Universitaria, Ciudad de M\'exico, M\'exico}
\affiliation{AI Factory at BBVA M\'exico}

\author{Francisco J. Sevilla}
\email[]{fjsevilla@fisica.unam.mx}
\affiliation{Instituto de F\'isica, Universidad Nacional Aut\'onoma de M\'exico,
Apdo.\ Postal 20-364, 01000, Ciudad de M\'exico, M\'exico}


\begin{abstract}
We analyze \emph{fractional Brownian motion} and \emph{scaled Brownian motion} on the two-dimensional sphere $\mathbb{S}^{2}$. 
We find that the intrinsic long time correlations that characterize fractional Brownian motion collude with the specific dynamics (\emph{navigation strategies}) carried out on the surface giving rise to rich transport properties. We focus our study on two classes of navigation strategies: one induced by a specific set of coordinates chosen for $\mathbb{S}^2$ (we have chosen the spherical ones in the present analysis), for which we find that contrary to what occurs in the absence of such long-time correlations,
\emph{non-equilibrium stationary distributions} are attained. These results resemble those reported in confined flat spaces in one and two dimensions [Guggenberger {\it et al.} New J. Phys. 21 022002 (2019), Vojta {\it et al.} Phys. Rev. E 102, 032108 (2020)], however in the case analyzed here, there are no boundaries that affects the motion on the sphere.  
In contrast, when the navigation strategy chosen corresponds to a frame of reference moving with the particle (a Frenet-Serret reference system), then the \emph{equilibrium} \emph{distribution} on the sphere is recovered in the long-time limit. For both navigation strategies, the relaxation times towards the stationary distribution depend on the particular value of the Hurst parameter. We also show that on $\mathbb{S}^{2}$, scaled Brownian motion, distinguished by a time-dependent diffusion coefficient with a power-scaling, is independent of the navigation strategy finding a good agreement between the analytical calculations obtained from the solution of a time-dependent diffusion equation on $\mathbb{S}^{2}$, and the numerical results obtained from our numerical method to generate ensemble of trajectories.
\end{abstract}


\maketitle

The motion on non-Euclidean surfaces has gained great interest, particularly on the surface of a two-dimensional sphere $\mathbb{S}^{2}$, where the collusion between the intrinsic dynamics of the moving objects and the curved surface they are moving on, leads to singular and interesting effects \cite{WisdomScience2003,Avron_NJPhys2006,SaportaKatzPRL2019,LinSoftMatter2020,Li_PNAS2022}. This growing interest is also encouraged by the fact that a variety of stochastic processes are equivalent to the diffusion of a tracer on the surface of a sphere, for instance, the tip of a unit vector that randomly rotates may describe the dynamics of a classical spin or the self-propulsion orientation of an active particle that moves with constant speed. Besides, the diffusive motion on curved surfaces is ubiquitous in different biological processes such as the motion of proteins or phospholipids (PIP2 \cite{FujiwaraJCellBio2002,MetzlerBiochBiophysActa2016}) on the cell membrane, which is crucial for cell signaling \cite{Haastert.2004,Ma.2004,Meer.2008,Meyers.2006,Hancock.2010}; cell migration confined to curved surfaces \cite{WernerAdvSci2017,BadeBioPhysJ2018,PieuchotNatureComm2018,LinSoftMatter2020}; or the pattern formation of epithelial tissue \cite{XiNatureComm2017,YuBiomaterials2021,Happel2022}. Additionally, the motion of \emph{active} particles diffusing on the sphere has been analyzed at the single particle and collective level \cite{CastroPRE2018,PraetoriusPRE2018,SknepnekPRE2015,ApazaPRE2017,ShankarPRX2017,Bao-quanSoftMatter2020,HenkesPRE2018,GrindovecPRB2020,HsuNatureComm2022,PiroNJPhys2022}.

An experimental analysis of the effects on particle diffusion due to the curvature of the underlying surface, has been carried out on polystyrene nanoparticles diffusing on the interface of a silicone-oil droplet immersed in water \cite{ZhongJPhysChemC}, while theoretical analysis of different effects---curvature gradient, external forces, many-body interactions---on the diffusion of particles confined to surfaces have been analyzed lately \cite{RamirezGarzaPhysChemChemPhys2021,ValdesJStatMech2021,Montanez-RodriguezPhysicaA2021,Ledesma-DuranFrontiersPhys2021}.

Despite of the important advances just mentioned, to our knowledge, the effects of long-time correlations on the motion of a particle moving on the surface of a compact manifold has not been studied, yet being of relevance since such processes model diverse mechanisms that lead to persistent (and antipersistent as well) motion on curved surfaces. In this article we address these aspects and fill this gap by analyzing the effects of long-time correlated diffusion on the one hand, and the effects of a time-dependent diffusion coefficient on the other, of a tracer particle moving on the surface of the two-dimensional sphere, $\mathbb{S}^{2}$. These two processes may arise from the complex interactions between the diffusing particle and the variety of components that lie on the sphere surface, such as in the cell membrane. A relevant aspect of our study is the elucidation of the nontrivial concomitant effects caused by the long-time correlations of the diffusive process of a tracer particle on a compact manifold, $\mathbb{S}^{2}$ in this case, and distinct navigation strategies.

The observation of long-time correlated Brownian processes, i.e., those for which the autocorrelation function of the particles position decays slowly with time in the form of a power law time-dependence, has been recognized to be ubiquitous in nature. This kind of correlations generally occurs in Brownian-like motion of particles in crowded environments, where the complex coupling of the Brownian particle with the environment leads to long-lasting correlations. This has been pointed out since the classical work of Alder and Wainwright \cite{AlderPhysRevLett1967,AlderPhysRevA1970} and of Widom \cite{WidomPhysRevA1971}, where it is shown that the coupling of a hard-core Brownian particle moving in an incompressible viscous fluid, leads to a velocity autocorrelation function that decays asymptotically as $t^{-3/2}$, instead of the exponentially fast relaxation, thus pointing out the role of the environment through hydrodynamic interactions \cite{PomeauPhysRep1975}. Following this rationale, a modification of the Langevin equation is carried out to take into consideration the fluid inertia \cite{HinchJFluidMech1975} and recovering the algebraic decay $t^{-3/2}$ of the velocity correlation function. Algebraic decay was subsequently observed in the experiments of Paul and Pusey \cite{PaulJPhysA1981} and Clercx and Schram \cite{ClercxPhysRevA1992}.

The overdamped motion of a particle moving in complex crowded environments, can exhibit a variety of behaviors. The most interesting being perhaps, \emph{anomalous diffusion}, which is characterized by the power-law time scaling $t^{\alpha}$ found for the time dependence of the
mean squared displacement, $\langle x^{2}\rangle\sim t^{\alpha}$, with $\alpha>0$, but $\alpha\neq1$.
It is widely accepted that such scaling is ubiquitous and originates from the long-time correlations of motion, which in turn, depend on the particular coupling between the environment and the particle \cite{HoflingRepProgPhys2013}. For instance, subdiffusion ($0<\alpha<1$) of submicron tracers was observed experimentally in the the motion of proteins embedded in the membranes of living cells \cite{WeissBioPhysJ2003,WeigelProcNatAcadSci2011,ManzoPhysRevX2015,HeNatComm2016}; also in the cytoplasm of biological cells \cite{WeissBioPhysJ2004,CaspiPhysRevLett2000,SeisenbergerScience2001,GoldingPhysRevLett2006,JeonPhysRevLett2011,TabeiProcNatAcadSci2013}; and in crowded liquids \cite{BanksBiophysJ2005,SzymanskyPhysRevLett2009,JeonNewJPhys2013}.

\begin{figure*}
 \begin{minipage}{0.49\textwidth}
	\includegraphics[width=\textwidth, trim= 0 0 0 0, clip]{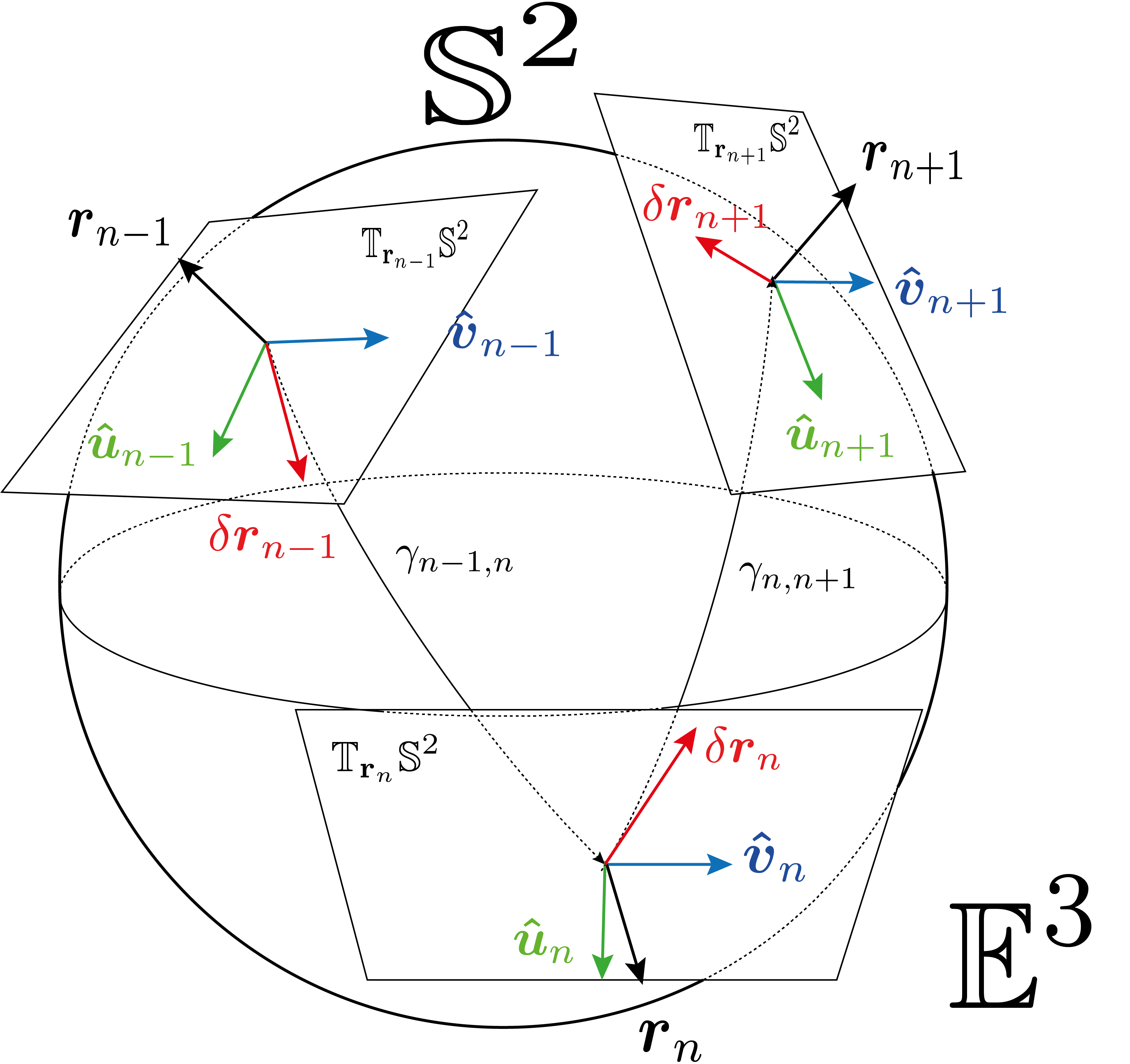}
	     \label{fig:subfig1}
	\end{minipage}
	\begin{minipage}{0.49\textwidth}
	    \includegraphics[width=0.32\textwidth, trim=0 25 0 25,clip=true]{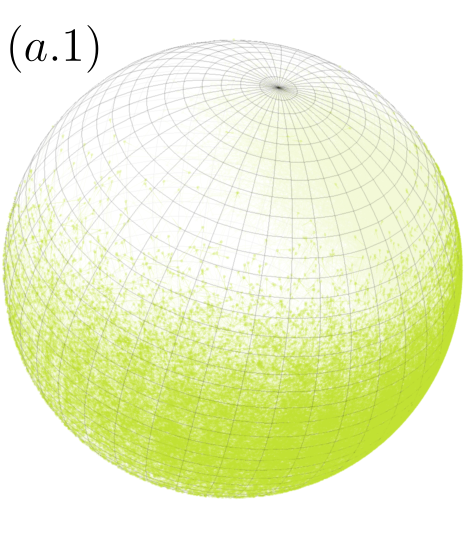}\,\includegraphics[width=0.32\textwidth,trim=0 20 0 25,clip=true]{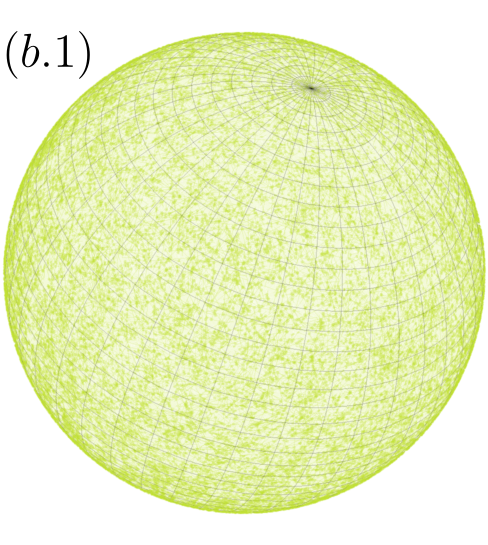}\,\includegraphics[width=0.32\textwidth]{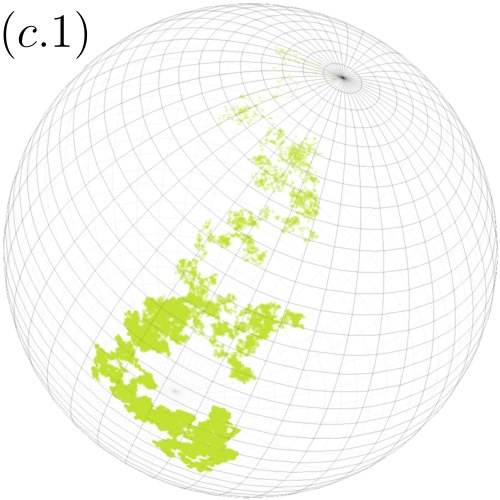}\\
\includegraphics[width=0.32\textwidth]{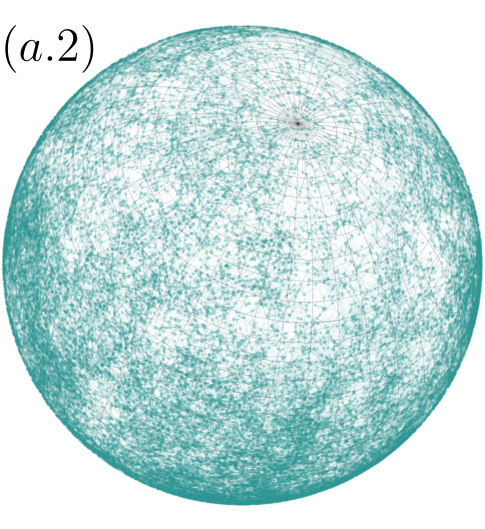}\,\includegraphics[width=0.32\textwidth]{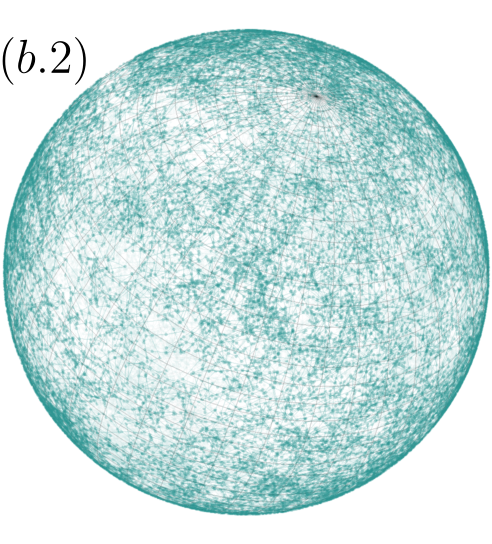}\,\includegraphics[width=0.32\textwidth]{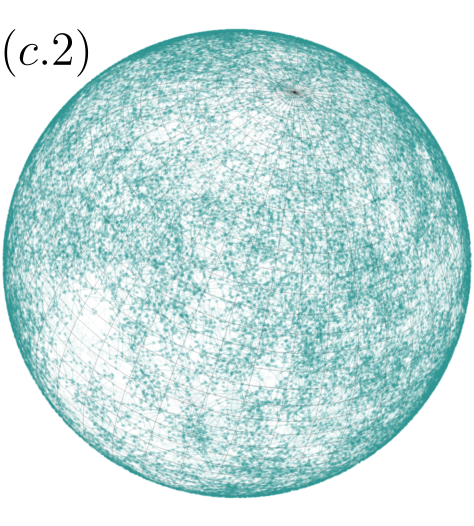}\\
\includegraphics[width=0.32\textwidth,trim=0 25 0 25,clip=true]{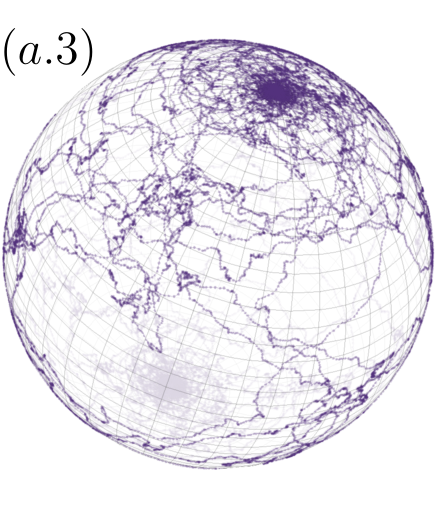}\,\includegraphics[width=0.32\textwidth,trim=0 0 0 0,clip=true]{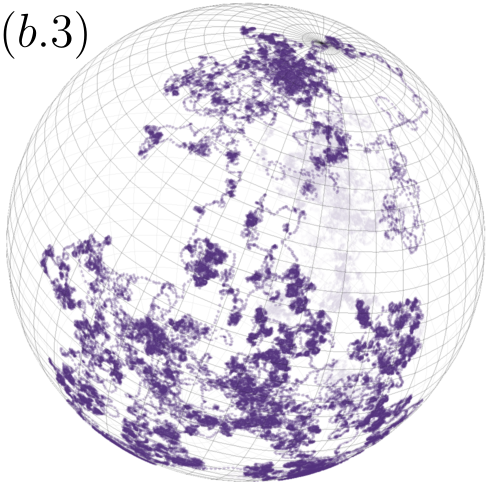}\,\includegraphics[width=0.32\textwidth]{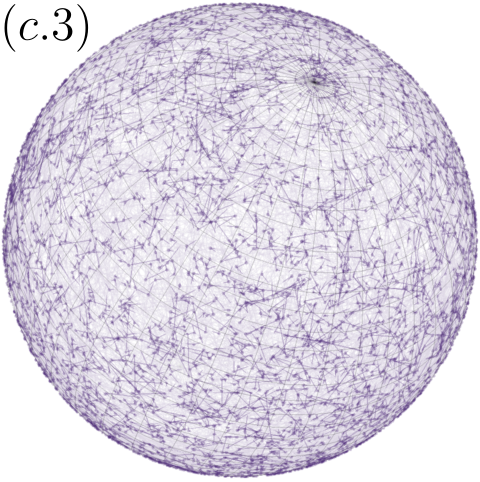}
	    \label{fig:subfig2}
        \end{minipage}
\caption{Left.- Schematic illustration of the numerical method used to generate trajectories of fractional and scaled Brownian motion on $\mathbb{S}^{2}$.
Two consecutive updating steps are shown indicating the transition from the $(n-1)$-th particle position, $\boldsymbol{r}_{n-1}$, to the $(n+1)$-th one $\boldsymbol{r}_{n+1}$, passing through $\boldsymbol{r}_n$ (depicted as normal vectors to their corresponding tangent planes $\mathbb{T}_{\boldsymbol{r}}\mathbb{S}^{2}$) as explained in the text. At each $\mathbb{T}_{\boldsymbol{r}}\mathbb{S}^{2}$, the basis vectors $\hat{\boldsymbol{u}}$'s, $\hat{\boldsymbol{v}}$'s, which define a particular navigation strategy, and the displacements  $\delta\boldsymbol{r}$'s, which are necessary to implement the updating rule \eqref{UpdatingRule}, are shown. Right.- Single particle trajectories on the sphere are shown. 
For fBm, trajectories were generated for distinct navigation strategies: Spherical system $\{\hat{\boldsymbol{\theta}},\hat{\boldsymbol{\phi}}\}$ (first column shows (a.1) for $H=0.1$, (a.2) for $H=0.5$ and (a.3) for $H=0.9$ ) and Frenet-Serret system $\{\hat{\boldsymbol{t}},\hat{\boldsymbol{n}}\}$ (middle column shows (b.1) for $H=0.1$, (b.2) for $H=0.5$ and (b.3) for $H=0.9$).
For sBm (third column), trajectories are insensitive to the navigation strategy, but they are affect by the specific time dependence of the diffusion coefficient  $D(t)=2HD_Ht^{2H-1}$. The cases $H=0.1$, $H=0.5$ and $H=0.9$ are shown in (c.1), (c.2) and (c.3) respectively. For $H=0.5$, (a.2), (b.2) and (c.2), all the cases give the same Brownian dynamics.}
\label{fig:GeometricMethod}
\end{figure*}

Different mathematical frameworks (models) have been ingeniously devised to describe anomalous diffusion \cite{MetzlerPhysChemChemPhys2014}, and two of them stand out as they take into consideration the \emph{memory effects} that arise, in most of the cases, from a reduced description focusing on the dynamics of the tracer particle dynamics from the rest of the elements that influence its motion. One corresponds to the \emph{generalized master equation} \cite{*[{Hola}] [{The reader should focus on sect IV.}] KenkreChapter1977} (or \emph{Continuous-Time Random Walks} \cite{KenkreJStatPhys1973}), which for power-law time memories leads to the \emph{fractional diffusion equations} of Metzler and Klafter and other authors \cite{WyssJMathPhys1986,MetzlerPhysReps2000}, that incorporate the fractional derivative, where the order of integrals and derivatives are generalized to fractional order \cite{PodlubnyBook,TarasovIntJModPhys2013,WestRevModPhys2014,SandevBook2019}. The other model that considers memory effects goes back to the seminal work of Kubo, and nowadays is referred as the \emph{generalized Langevin equation} \cite{Kubo_RepProgPhys1966}. This takes into account the retarded effect of the dissipative force and becomes the \emph{fractional Langevin equation} when the memory is chosen in the class of memory function with power-law time dependence \cite{KobelevProgTheorPhys2000,LutzPhysRevE2001,BurovPhysRevLett2008}. This theoretical framework has found diverse applications, particularly in the analysis of the diffusion of Brownian motion in viscoelastic fluids \cite{RodriguezJPhysA1988}, where the hydrodynamic interaction, as noticed by Boussinesq (see Ref. \cite{MakrisPhysFluids2021} for a historical recount) plays an important role in the dynamics.

In this work we resort to \emph{fractional Brownian motion} (fBm) as a simple model of long-time correlated motion which was introduced by Mandelbrot and van Ness (and based in the previous work by Kolgomorov, Hurst, Hunt, Lamperti and Yaglom), to model natural time series that exhibits extremely long interdependence \cite{MandelbrotSIAMRev1968}. This stochastic process has gained interest in different contexts, particularly regarding anomalous diffusion, due to its simple, but non-trivial, intrinsic characteristics: being Gaussian, self-similar and with stationary increments \cite{MetzlerPhysChemChemPhys2014,SokolovSoftMatter2012,BiaginiBook2008}. Although the ``fractional'' character in this case is different from the one of fractional derivatives, a connection with the overdamped limit of a fractional Langevin equation has been conveyed in Ref. \cite{SokolovSoftMatter2012}.

In addition, we also consider the case of \emph{scaled Brownian motion} (sBm), which defines a Gaussian process characterized by the time-dependent diffusion coefficient $D(t)=2HD_{H}\, t^{2H-1}$. In unbounded Euclidean space, sBm is defined by the solution of the  diffusion equation with time-dependent diffusion coefficient and leads to the power-law scaling of the mean squared displacement $\langle\boldsymbol{x}^{2}(t)\rangle\sim t^{2H}$ \cite{JeonPhysChemChemPhys2014,ThielPhysRevE2014}, although no long-time correlated motion is involved. In here we use the same symbols $H$ and $D_{H}$ as in fBm since they have the same physical units, however, they have different physical meaning as they characterize two different stochastic processes; in  Ref. \cite{JeonPhysChemChemPhys2014}, it is shown that under confinement, the stochastic process given by sBm is fundamentally distinct from fBm. We show in this paper that this is also the case when the motion occurs freely on $\mathbb{S}^2$.

We analyze a statistical ensemble of discretized trajectories generated by extending the method presented in Ref. \cite{ValdesJStatMech2021}. Such an extension takes into account either the long-time correlations of the motion, or the time-dependence of the diffusion coefficient of a non-stationary diffusion process. Briefly, an element of the ensemble is generated by recursively updating the particle position, initiating at $\boldsymbol{r}_0$ on $\mathbb{S}^2$ at time $t=0$, such that in a time interval $\Delta t$, the particle position at the $(n+1)$-th step is obtained from the previous one as 
\begin{equation}\label{UpdatingRule}
\boldsymbol{r}_{n+1} = \prod_{T \mathbb{S}^2_{\boldsymbol{r}_{n}} \to \mathbb{S}^2} \biggl[ \boldsymbol{r}_{n}+\alpha_{n}  \delta\boldsymbol{r}_{n} \biggr], \end{equation}
where $\delta\boldsymbol{r}_{n}=\delta U(\Delta t)\, \hat{\boldsymbol{u}}_{n}+\delta V(\Delta t)\, \hat{\boldsymbol{v}}_{n}$ lies on the tangent plane $\mathbb{T}_{\boldsymbol{r}_{n}}\mathbb{S}^2$ centered at $\boldsymbol{r}_{n}$ on the sphere; $\{\hat{\boldsymbol{u}}_{n},\hat{\boldsymbol{v}}_{n}\}$ is an orthonormal basis of  $\mathbb{T}_{\boldsymbol{r}_{n}}\mathbb{S}^2$ (see left panel in Fig. \ref{fig:GeometricMethod}); and $\delta U(\Delta t)$, $\delta V(\Delta t)$ are two statistically independent increments obtained from the underlying stochastic process considered. The updating rule \eqref{UpdatingRule} considers the factor
\begin{equation}\label{ScalingFactor}
\alpha_{n}=\frac{r}{\bigr|\bigr|\delta\boldsymbol{r}_{n}\bigr|\bigr|} \tan \biggl \{\frac{\bigr|\bigr|\delta\boldsymbol{r}_{n}\bigr|\bigr|}{r} \biggr\},    
\end{equation}
which rescales $\delta\boldsymbol{r}_{n}$ such that $\boldsymbol{r}_n+\alpha_n \delta\boldsymbol{r}_n$ is projected along the corresponding geodesic $\gamma_{n,n+1}$ with the correct arc length. The operator $\prod_{T \mathbb{S}^2_{\boldsymbol{r}} \to \mathbb{S}^2}[\boldsymbol{x}]$ being  $\hat{\boldsymbol{x}}r$, with $r$ the sphere radius (unit vectors are defined as $\hat{\boldsymbol{x}}=\boldsymbol{x}/||\boldsymbol{x}||$). The tangent plane and the four referred vectors are displayed in the left panel of Fig. \ref{fig:GeometricMethod}, for three consecutive iterations. The positions $\boldsymbol{r}_n$, orthogonal to $\mathbb{T}_{\boldsymbol{r}_{n}}\mathbb{S}^2$, are depicted in black, the displacements on the plane $\delta\boldsymbol{r}_{n}$ are depicted in red, while the basis vectors $\hat{\boldsymbol{u}}_{n}$ and $\hat{\boldsymbol{v}}_{n}$ are depicted in green and blue, respectively. 

The long-time correlated motion considered for our analysis is carried out by mapping to the sphere surface (through the use of Eq. $\eqref{UpdatingRule}$), two statistically independent fBm processes. In the case when the stochastic motion is characterized by a time-dependent diffusion coefficient, we use the so-called \emph{scaled Brownian motion} (sBm) for which the diffusion coefficient $D(t)$ scales with time as $t^{\beta}$ with $-1<\beta<1$. 

\paragraph*{Fractional Brownian motion on $\mathbb{S}^2$.-} To generate long-time correlated trajectories we consider two statistically independent fBm processes, $B_{1}^{H}(t)$ and $B_{2}^{H^\prime}(t)$, characterized by Hurst exponents $0<H,H^\prime<1$, respectively. These are Gaussian stochastic processes with $\langle B^H(t)\rangle=0$ and autocorrelation function
\begin{equation}\label{fBM-Correlation}
\bigl\langle B^{H}(t)B^{H}(s)\bigr\rangle=D_{H}\bigl(t^{2H}+s^{2H}-\vert t-s\vert^{2H}\bigr),
\end{equation}
where $D_{H}$ (with units of [length]$^{2}\times$[time]$^{-2H}$) measures its amplitude. Eq. \eqref{fBM-Correlation} reduces to the auto-correlation of the Wiener process $2D_{1/2}\textrm{min}(t,s)$ when $H=\frac{1}{2}$ and gives $\langle[B^H(t)]^2\rangle=D_H\, t^{2H}$ at equal times for all $H$ in $(0,1)$. The fBm process is also defined as the integral of \emph{fractional Gaussian noise} $\xi^H(t)$, as $B^H(t)=\int_{0}^tds\, \xi^{H}(s)$, $\xi^H(t)$ is a stationary stochastic process with autocorrelation function $\langle\xi(t)\xi(0)\rangle=2HD_Ht^{2H-1}[(2H-1)t^{-1}+\delta(t)]$ \cite{Qian2003}, from which is recognized a time-dependent diffusion coefficient $D(t)=2HD_Ht^{2H-1}$.

We are interested in the case for which $H=H^{\prime}$, however an extension to the anisotropic case is straightforward. The increments of the fBms during the time interval $\Delta t=t-t^{\prime}$, $t>t^{\prime}$, $\delta B_{1}^{H}(\Delta t)\equiv B_{1}^{H}(t)-B_{1}^{H}(t^{\prime})$ and $\delta B_{2}^{H}(\Delta t)\equiv B_{2}^{H}(t)-B_{2}^{H}(t^{\prime})$ (corresponding to the increments $\delta U(\Delta t)$ and $\delta V(\Delta t)$ respectively),
are stationary, long-time correlated, and not independent if $H\neq 0.5$. In the case $0<H<\frac{1}{2}$ they are negatively correlated, while they are positively correlated if $\frac{1}{2}<H<1$ \cite{BiaginiBook2008}. 
In this case the increment $\delta\boldsymbol{r}_{n}$ is given by $\delta B_{1}^{H}(\Delta t)\, \hat{\boldsymbol{u}}_{n}+\delta B_{2}^{H}(\Delta t)\, \hat{\boldsymbol{v}}_{n}$, with $\{\hat{\boldsymbol{u}}_n,\hat{\boldsymbol{v}}_n\}$ an orthonormal vector basis for $\mathbb{T}_{\boldsymbol{r}_{n}}\mathbb{S}^2$. 

We found that the long-time correlations of fBm ($H\neq\frac{1}{2}$) collude with the specific implementation of the updating rule \eqref{UpdatingRule} through the choice of the orthonormal basis $\{\hat{\boldsymbol{u}}_n,\hat{\boldsymbol{v}}_n\}$, which defines a kind of \emph{navigation strategy} that drives the particle motion to give rise to a rich variety of statistical patterns of motion on the sphere. On the contrary, uncorrelated motion (as is the standard Brownian motion, $H=1/2$) is insensitive to the choice of such strategies [see (a.2) and (b.2) for a qualitative comparison in the right panel in Fig. \ref{fig:GeometricMethod}]. 

We consider for our analysis two contrasting  physically-motivated navigation strategies, one is observed in many biological organisms which may have a distinct body axis (head-tail axis) defining a preferred direction of motion  sometimes referred as \emph{heading}. In such a case the particle heading $\hat{\boldsymbol{t}}_n$, together with the normal vector $\hat{\boldsymbol{n}}_n$, form the orthonormal basis for the updating rule \eqref{UpdatingRule} that carry the long-term correlated motion. Notice that with $\hat{\boldsymbol{r}}_n$ (the binormal vector), these three vectors form the well-known Frenet-Serret system. Three sample trajectories are shown: (b.1), (b.2) and (b.3), in the second column of the right panel of Fig.~\ref{fig:GeometricMethod}, for highly anticorrelated motion, $H=0.1$ (b.1), uncorrelated motion, $H=0.5$ (b.2), and highly correlated motion, $H=0.9$ (b.3).

The other navigation strategy corresponds to the case for which the orthonormal basis for the updating rule, $\{\hat{\boldsymbol{\theta}}_n,\hat{\boldsymbol{\phi}}_n\}$, forms a kind of ``laboratory reference frame'', which for our analysis is chosen to be the one induced by the spherical coordinates of $\mathbb{S}^2$, and consequently attached to each point of the sphere surface, namely, $\hat{\boldsymbol{\theta}}_{n}=(\cos\theta_n\cos\phi_n,\cos\theta_n\sin\phi_n,-\sin\theta_n)$, oriented along the sphere meridians, and $\hat{\boldsymbol{\phi}}_n=$ $(-\sin\theta_n$ $ \sin\phi_n,$ $\sin\theta_n\cos\phi_n,0)$ orthogonally oriented along the parallels (see left panel in Fig.~\ref{fig:GeometricMethod}). As a consequence, the long-time correlated trajectories describe a distinct pattern as is shown by the trajectories shown in the right panel of Fig.~\ref{fig:GeometricMethod} for $H=0.1$ (a.1) and $H=0.9$ (a.3). \emph{Self-confinement} is observed for  $H\neq0.5$, each particle excursion is limited to a sector of the sphere around well-defined time-average values $\overline{\theta}$, $\overline{\phi}$ that depend on $H$. For $0<H<0.5$, particles concentrate forming an island around $\overline{\theta}=\pi/2$ (equator) {with} $\overline{\phi}$ {randomly} chosen; while for $0.5 < H < 1$ distinctively, the particles concentrate their excursions around the poles, i.e. $\overline{\theta}=0,\pi$, wandering between them as is shown in Fig.~\ref{fig:GeometricMethod}(a.3). For this navigation strategy, ensemble averages strikingly differs from time averages giving rise to weak ergodicity breaking \cite{ThielPhysRevE2014}, especially for $0<H<0.5$, for which single trajectories get trapped diffusing around a randomly chosen sector of the sphere, thus, the time average using a single long trajectory will differ from the ensemble average which samples uniformly those sectors; the equivalence between ensemble averages and time averages is recovered for $H=1/2$ [see a trajectory generated in Fig.~\ref{fig:GeometricMethod}(a.2)].

\paragraph*{Scaled Brownian motion on $\mathbb{S}^{2}$.-} The other pattern of stochastic motion we address in this paper refers to sBm. The updating rule \eqref{UpdatingRule} in this case considers two independent Gaussian processes $G_{1}(t)$ and $G_{2}(t)$, with statistically independent increments $\delta G_{1}(\Delta t_{n})=\sqrt{2{D_H\bigl[{t_n}^{2H}-{t_{n-1}}^{2H}\bigr]}}\, \Delta W_1$, and  $\delta G_{2}(\Delta t_n)=\sqrt{2{D_H\bigl[{t_n}^{2H}-{t_{n-1}}^{2H}\bigr]}}\, \Delta W_2$, for the time increment $\Delta t_n=t_n-t_{n-1}$, $n=1,\,2\,\ldots$; $\Delta W_1$, $\Delta W_2$ are two independent Wiener processes of vanishing mean and unitary variance{; and $t_{n}=\sum_{k=1}^{n}\Delta t_{k}$ being the total time elapsed up to step $n$}. Scaled Brownian motion is highly non-stationary due to the time-dependence of the diffusion coefficient, a feature that is implemented by considering {$\delta\boldsymbol{r}_{n}=\sqrt{2D_{H}{\Delta t}^{2H}\bigl[n^{2H}-(n-1)^{2H}\bigr]}\Bigl[\Delta W_{1}\, \hat{\boldsymbol{u}}_{n}+\Delta W_{2}\, \hat{\boldsymbol{v}}_{n}\Bigr]$}, {where we have assumed homogeneous time increments $\Delta t$}, and the pair of orthonormal vectors  $\{\hat{\boldsymbol{u}}_n,\hat{\boldsymbol{v}}_n\}$ is an arbitrary vector basis for $\mathbb{T}_{\boldsymbol{r}_{n}}\mathbb{S}^2$. It is the statistical independence among the increments what guaranties that the updating rule is independent of the basis choice, as occurs for {uncorrelated} standard Brownian motion (or fBm with $H=0.5$). Three sBm trajectories are shown in the third column of the right panel of Fig. \ref{fig:GeometricMethod} for $H=0.1$ (c.1), $H=0.5$ (c.2) and $H=0.9$ (c.3), which are contrasted with {those for the two navigation strategies considered for the fractional} Brownian motion case (trajectories in the first and second columns in the same figure).

The corresponding modification of the Euclidean Fokker-Planck equation that considers that sBm occurs on $\mathbb{S}^2$ is
\begin{multline}\label{SmoluOp}
\frac{\partial}{\partial t}P_\textrm{sBm}(\theta,\phi,t) = \frac{2HD_{H}}{r^{2}}t^{2H-1}\Bigl[\frac{1}{\sin\theta}\frac{\partial}{\partial\theta}\Bigl(\sin\theta\frac{\partial}{\partial\theta}\Bigr)+\\
\frac{1}{\sin^{2}\theta}\frac{\partial^{2}}{\partial\phi^{2}}\Bigr]P_\textrm{sBm}(\theta,\phi,t),
\end{multline}
where $\boldsymbol{r}=(r \sin{\theta} \cos{\phi}, r \sin{\theta} \sin{\phi}, r \cos{\theta})$ denotes the particle's position in spherical coordinates, $r$ being the radius of $\mathbb{S}^2$.  
The solution of \eqref{SmoluOp} is given by
\begin{equation}
P_\textrm{sBm}(\theta,\phi,t)=\sum_{l=0}^{\infty}\sum_{m=-l}^{l}p_{l}^{m}\, Y_{l}^{m}(\theta,\phi)
e^{-l(l+1)D_{H}t^{2H}/r^{2}},
\end{equation}
the coefficients $p_{l}^{m}$ are determined from the initial distribution, and $Y_{l}^{m}(\theta,\phi)$ are the standard spherical harmonics. If the initial condition corresponds to a localized distribution at the north pole we get
\begin{equation}\label{PDF}
P_\textrm{sBm}(\theta,\phi,t)=\sum_{l=0}^{\infty}\frac{2l+1}{4\pi}\, \text{P}_{l}(\cos\theta)\\
e^{-l(l+1)D_{H}t^{2H}/r^{2}},
\end{equation}
with $\textrm{P}_n(\cos\theta)$ the Legendre polynomials. The absence of $\phi$ in \eqref{PDF} implies the azimuthal symmetry of the process, which leads to the moments of the polar angle $\theta$,  $\langle\theta^{n}\rangle_\textrm{sBm}=\int d\Omega\, \theta^{n}P_\textrm{sBm}(\theta,\phi,t)$ given by
\begin{equation}
\langle\theta^{n}\rangle_\textrm{sBm}=\sum_{l=0}^{\infty}(2l + 1) g_{\theta^n}(l) e^{- l(l+1) D_{H}t^{2H}/r^2},
\end{equation}
where
$g_{\theta^n}(l)=\int_{0}^{\pi}d\theta\, \theta^{n}\, (\sin\theta/2)\,  \textrm{P}_{l}(\cos\theta)$. It is clear from \eqref{PDF} that for each $H$, the marginal stationary distribution of the polar angle, $P_{eq}(\theta)=\sin\theta/2$, is attained in the long-time regime, indicating the uniform distribution of the particle positions on $\mathbb{S}^2$.

In addition the auto-correlation function of the particle positions can be computed straightforwardly and is given by \begin{align}\label{ACFfBm}
   \langle \hat{\boldsymbol{r}}(t) \cdot \hat{\boldsymbol{r}}(0) \rangle_\text{sBm} = \exp{\left\{-2 D_{H}t^{2H}/r^2 \right \}}.
\end{align}
{Notice that the time dependence of the quantities computed from Eq. \eqref{SmoluOp}, namely $\langle\theta^n\rangle_\textrm{sBm}$ and $\langle\hat{\boldsymbol{r}}(t)\cdot\hat{\boldsymbol{r}}(0)\rangle_\textrm{sBm}$, become $H$-invariant under the time-scaling transformation $d\tau_H=2HD_H\, t^{2H-1}\, dt$. Such transformation makes Eq. \eqref{SmoluOp} $H$-invariant leading to dynamical scaling as can be checked straightforwardly.}

\begin{figure*}
\centering
\includegraphics[width=0.4\textwidth,trim=110 130 40 140,clip=true]{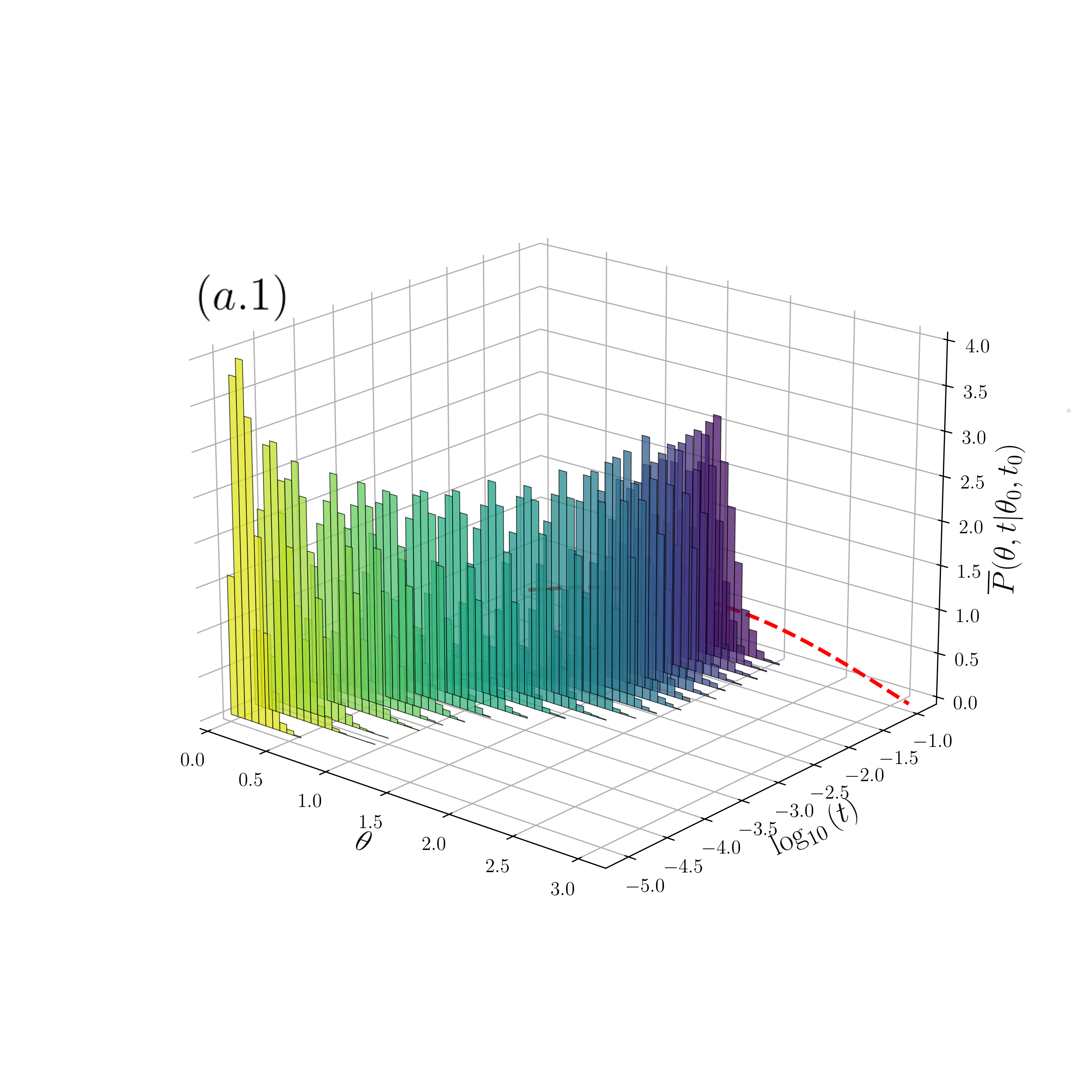}\qquad\includegraphics[width=0.4\textwidth,trim=100 130 45 150,clip=true]{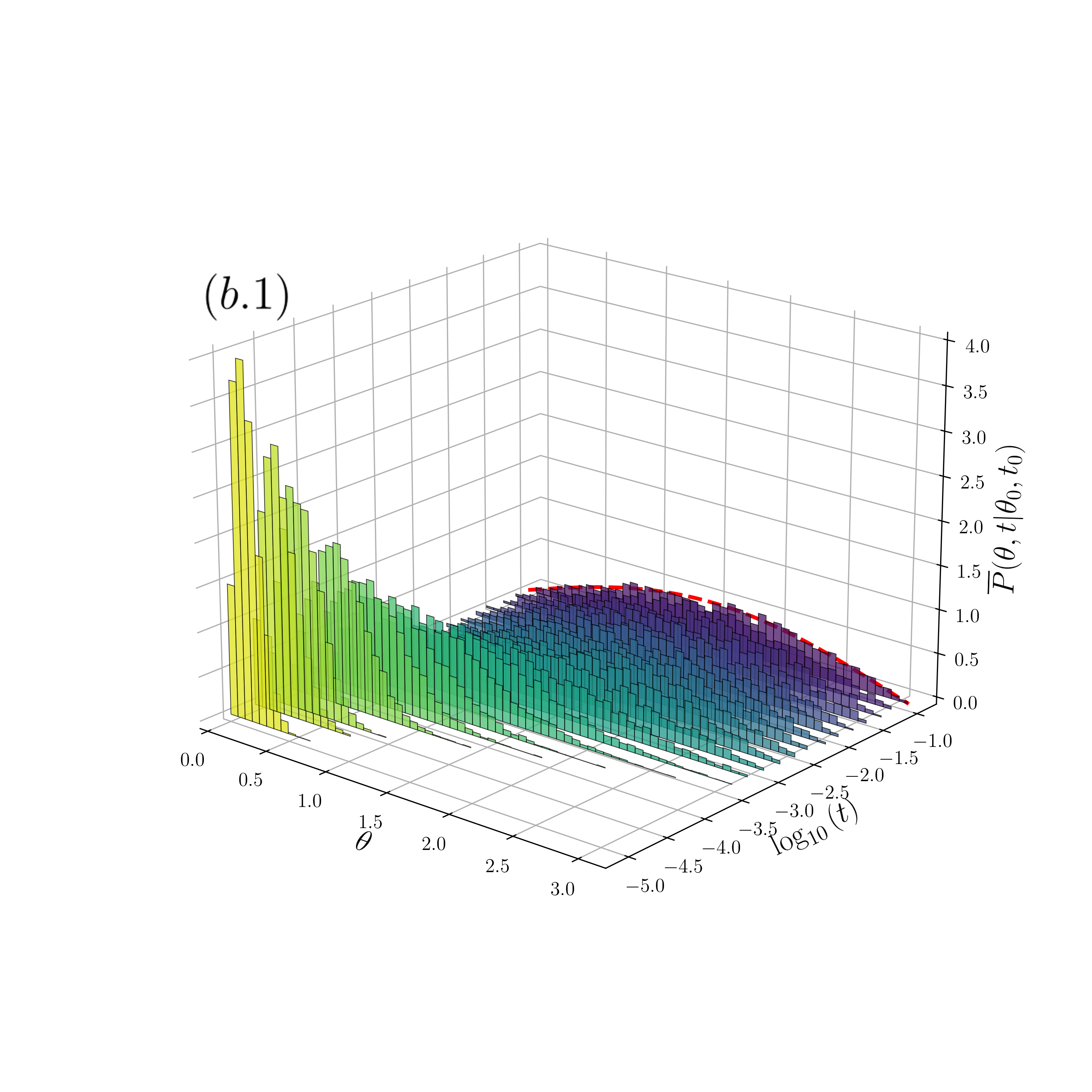}\\
\includegraphics[width=0.36\textwidth,trim=70 80 20 100,clip=true]{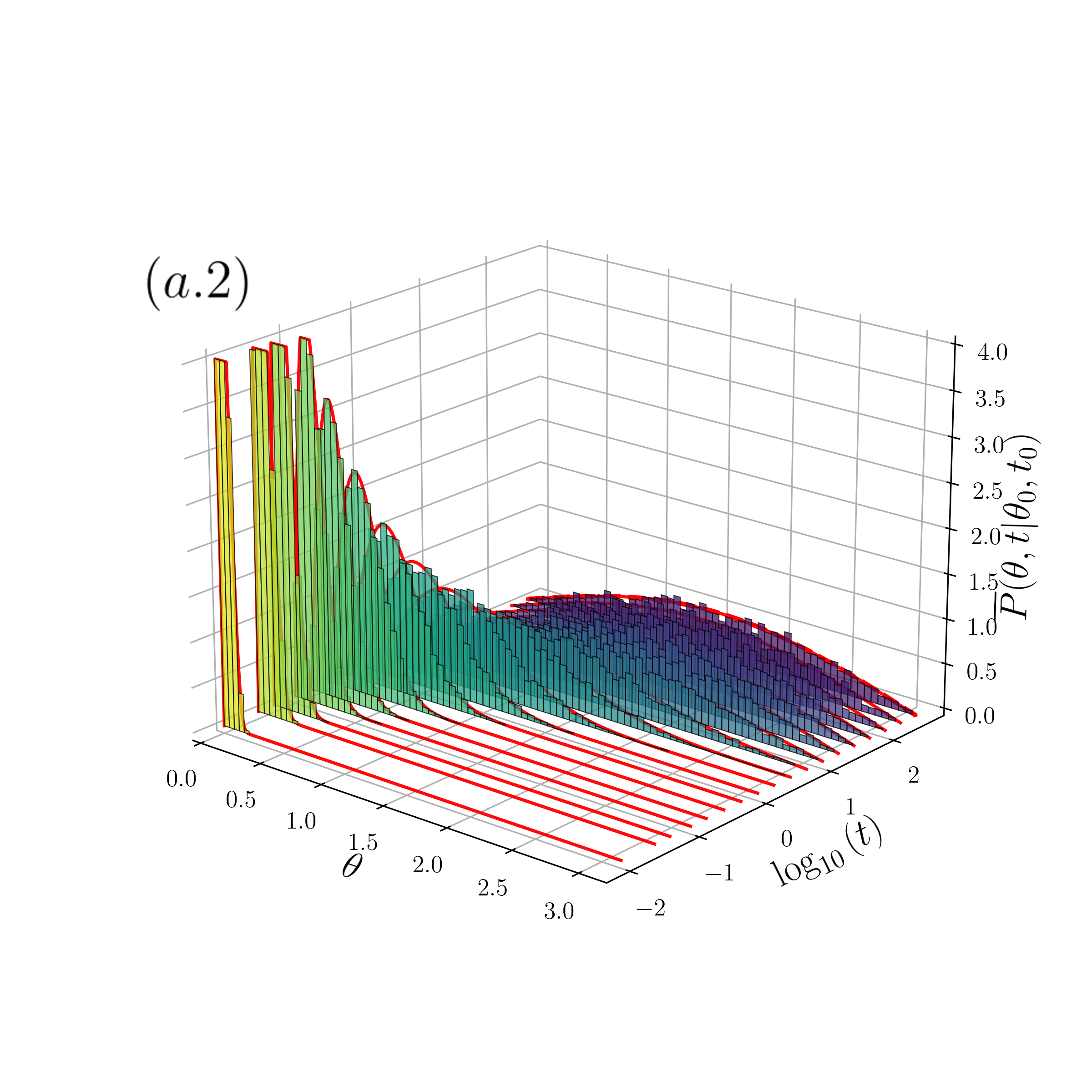}\qquad\qquad\includegraphics[width=0.4\textwidth,trim=100 130 45 150,clip=true]{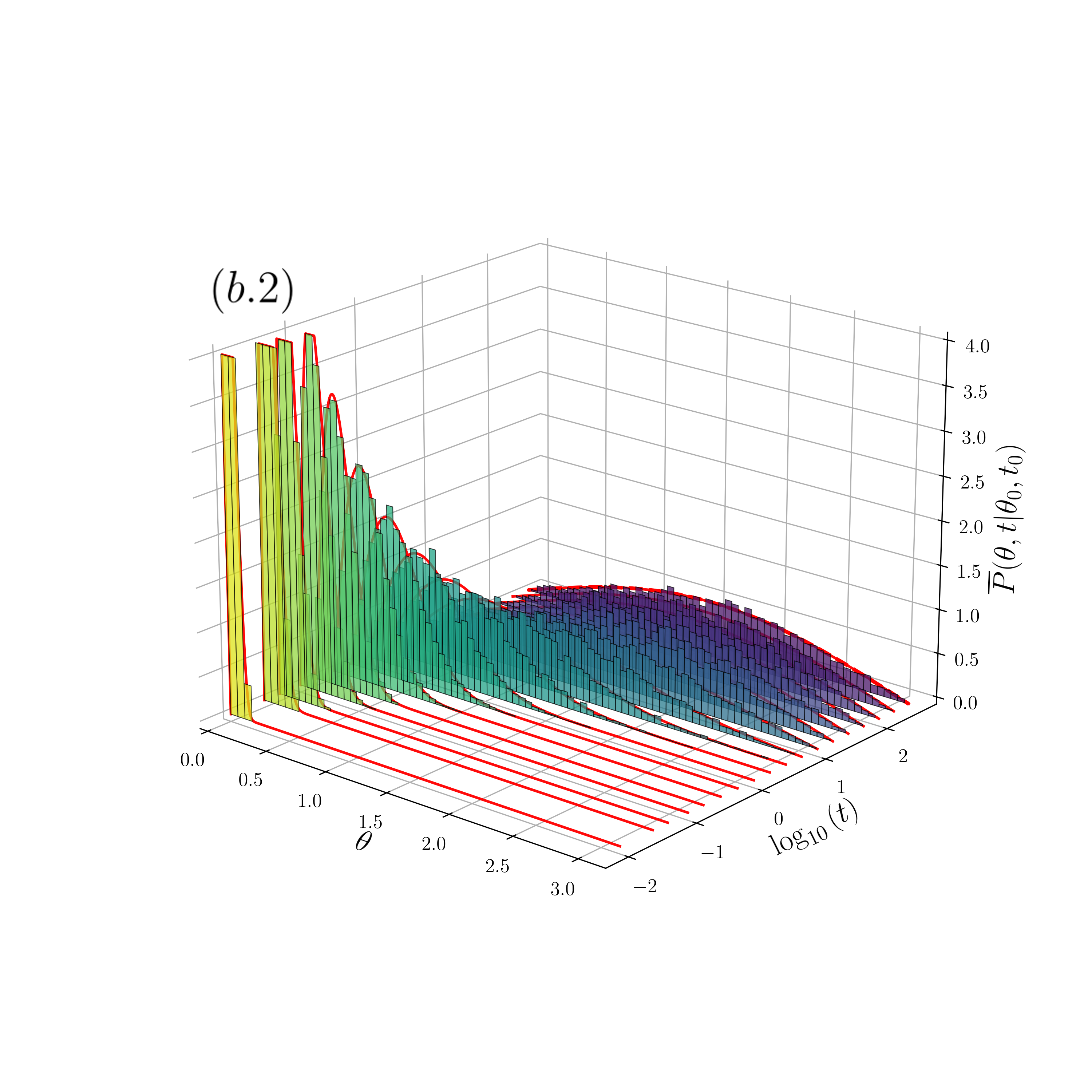}\\
\includegraphics[width=0.4\textwidth,trim=100 130 45 150,clip=true]{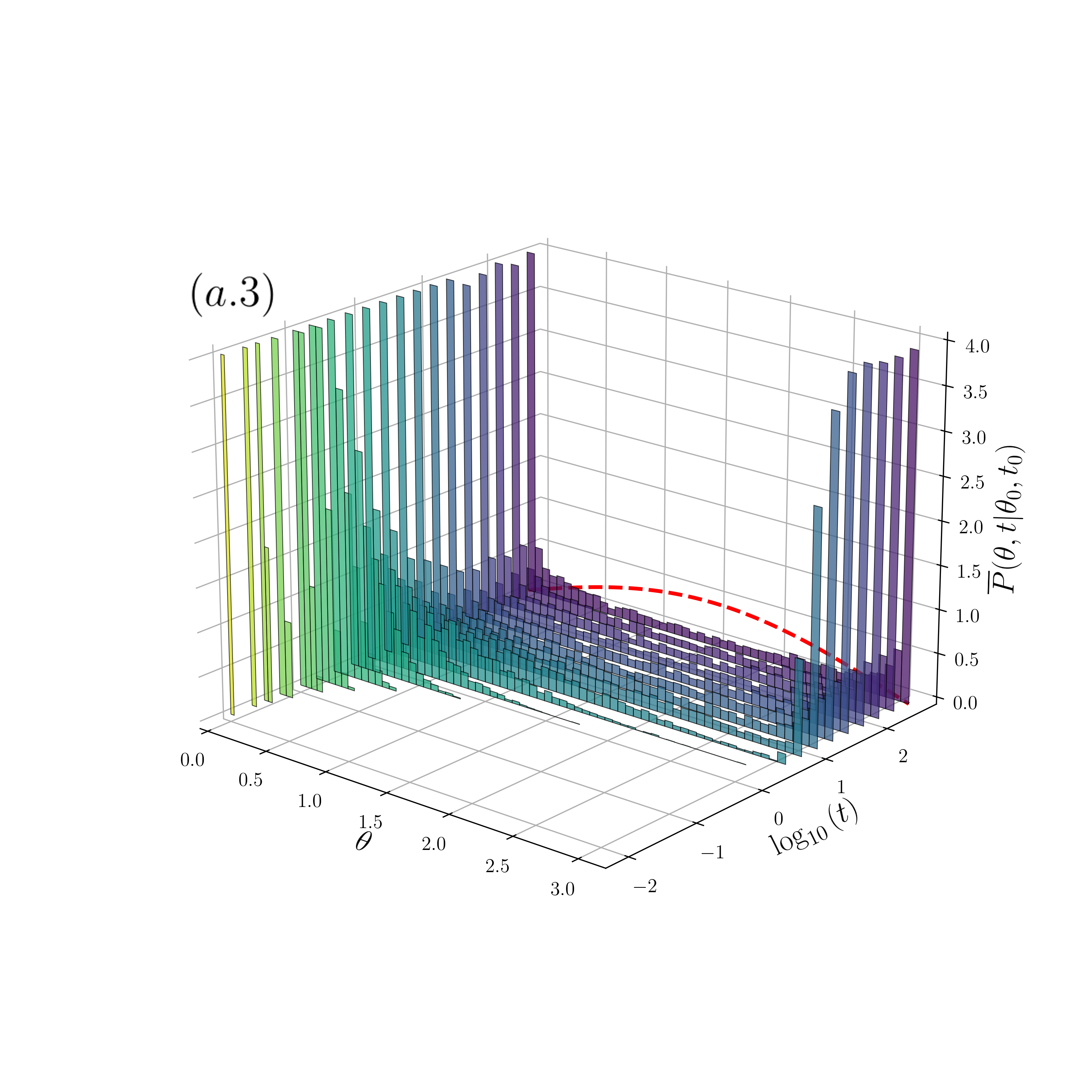}\qquad\includegraphics[width=0.4\textwidth,trim=100 130 45 150,clip=true]{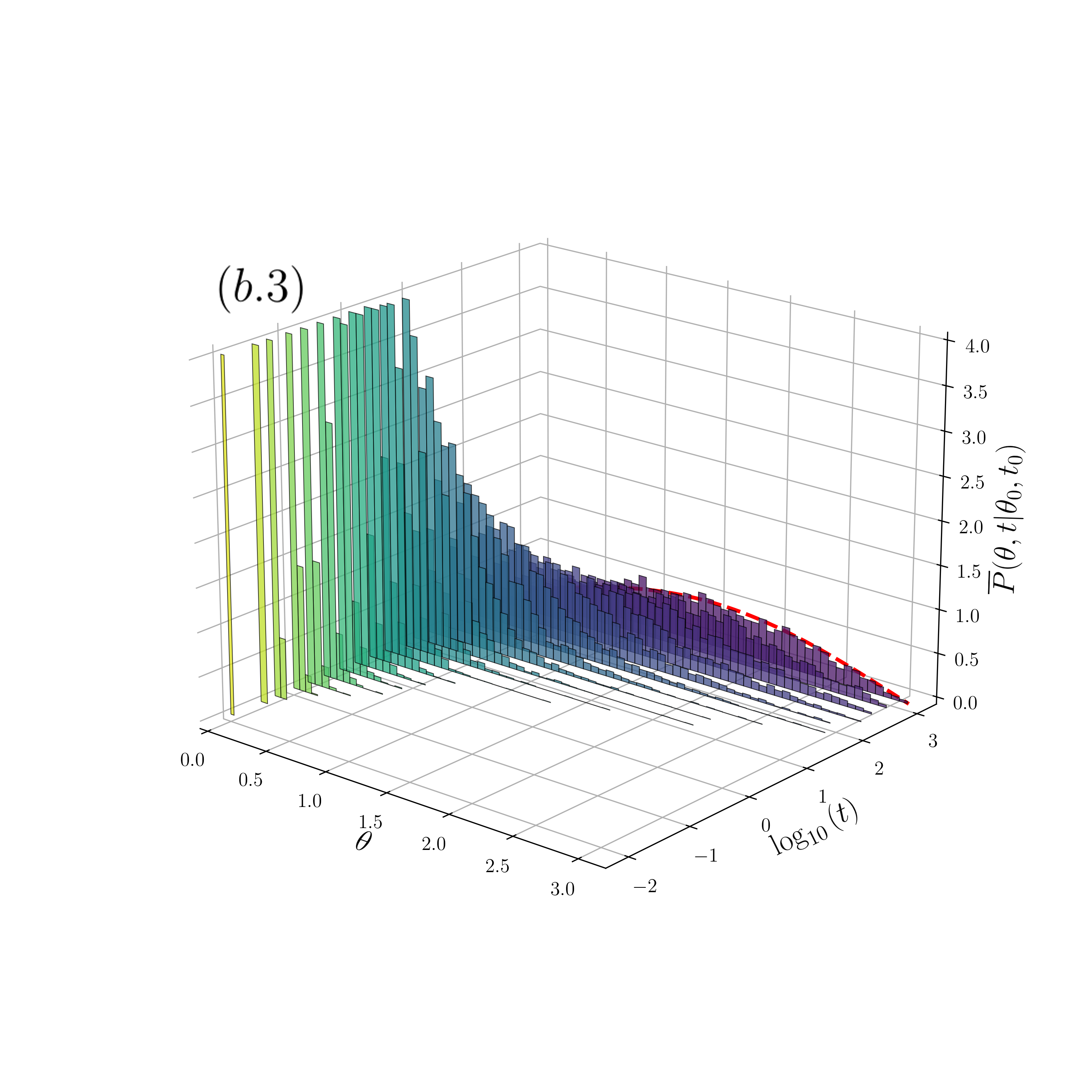}\\
\caption{The frequency distributions of the polar angle $\overline{P}(\theta,t\vert\theta_0,t_0)$ at different times ($\log t$) is shown for  
fBm in the spherical coordinates basis $\{\hat{\boldsymbol{\theta}},\hat{\boldsymbol{\phi}}\}$ [column (a)], 
fBm in the Frenet-Serret basis $\{\hat{\boldsymbol{t}},\hat{\boldsymbol{n}}\}$ [(b) column].  
Numerals 1, 2 and 3 correspond to the values $H=0.1$, $H=0.5$ and $H=0.9$, respectively. In all cases we have set $\theta_0=0$ at $t_0=0$.
For uncorrelated motion, $H=0.5$ in the [(a.2) and (b.2)], the stochastic dynamics is independent of the basis system chosen recovering Brownian motion on $\mathbb{S}^2$ (continuous-red lines depict the known analytical solution \cite{ValdesJStatMech2021}). 
In each graph the equilibrium stationary distribution of the polar angle, $P_{st}(\theta)=\sin\theta/2$ (uniform on the sphere), is displayed on the background with a red-dashed line. When the basis induced by spherical coordinates is used [(a) column] the distributions reach nonequilibrium stationary distributions, unimodal with postive excess kurtosisi respect to $P_{st}(\theta)$ for $0<H<0.5$ (shown only $H=0.1$ which is the most conspicuous nonequilibrium distribution), and bimodal with modes at the poles for $0.5<H<1$.}
\label{Theta-Time-Hist}
\end{figure*}

\paragraph*{Frequency distribution of the polar angle $\theta$ for fBm: The role of the navigation strategy.-} 
We present an statistical analysis of a set of ensembles of trajectories generated with the method described in the previous paragraphs: Two subsets corresponding to fBm, one for each of the navigation strategies considered{;} and one subset {corresponding} to sBm. {In the case of fBm} {e}ach ensemble consisted of 5$\times10^{3}$ trajectories for each of the nine values of the $H$ considered, namely $H=0.1,0.2,\ldots,0.9$ for which we have set $D_{H} = 0.1$ in arbitrary units. {For sBm the absence of long-time correlations allow to consider ensembles of 1.2$\times10^{5}$ trajectories.} 

\begin{figure}
\centering
\includegraphics[width=0.9\columnwidth]{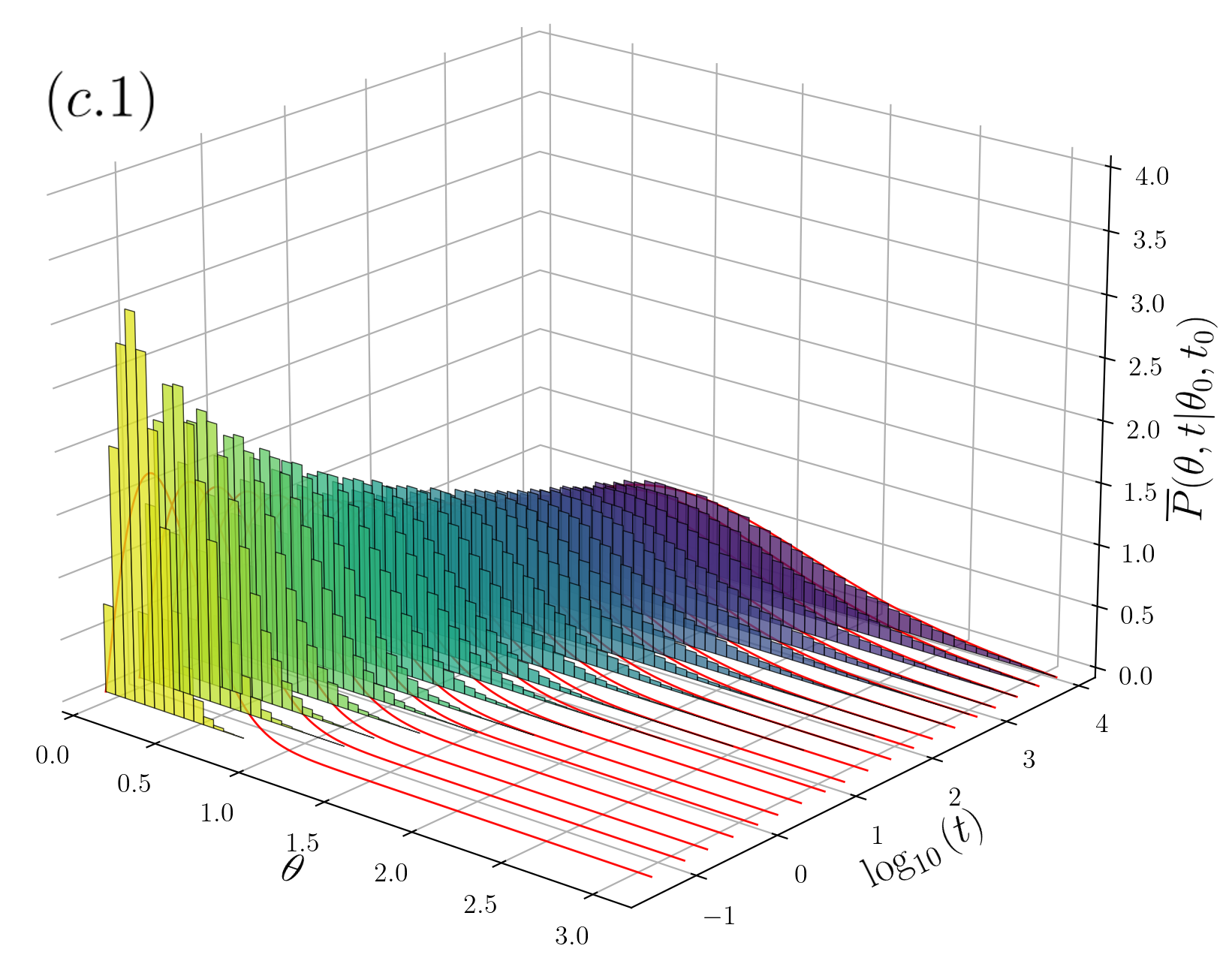}
\includegraphics[width=0.9\columnwidth]{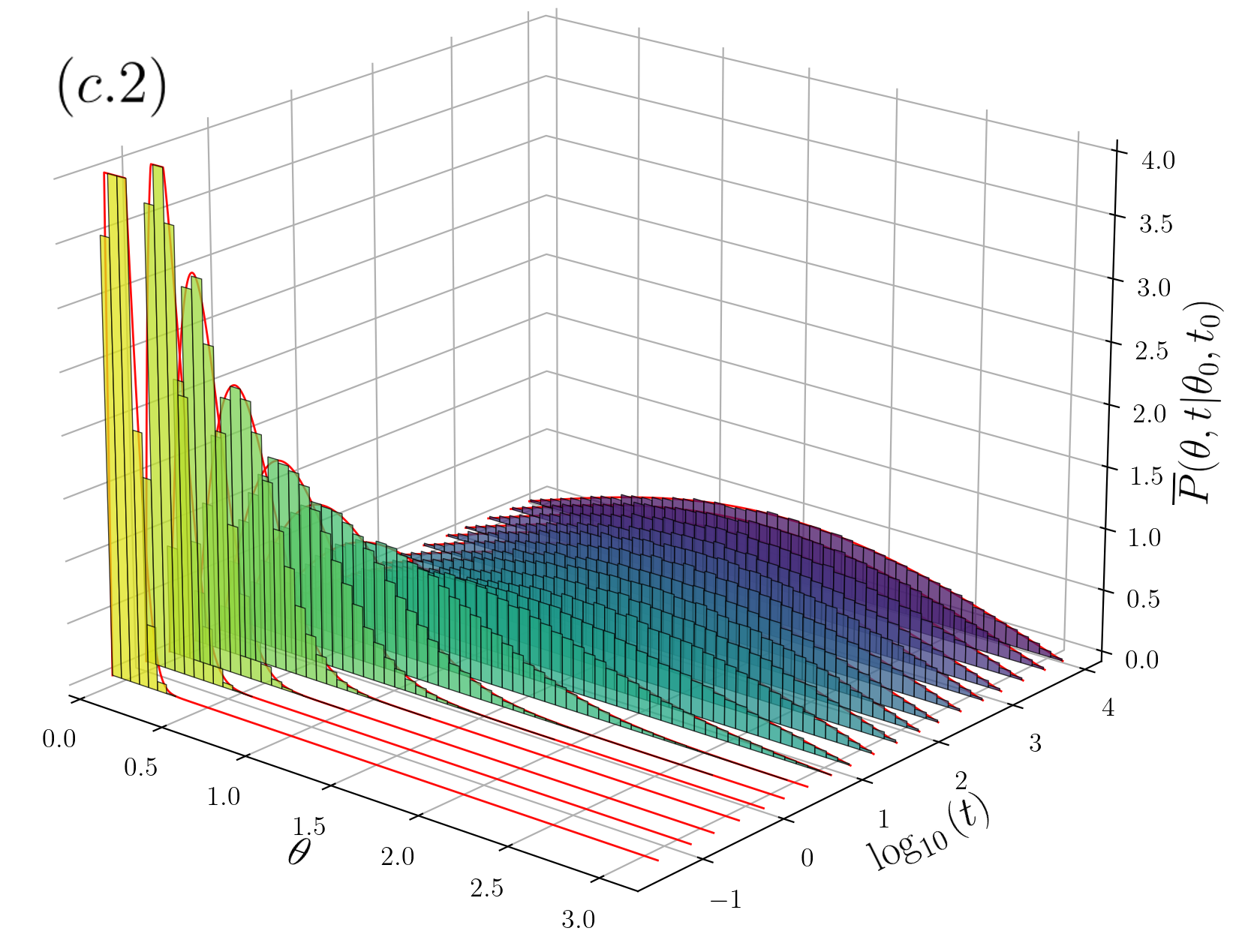}
\includegraphics[width=0.9\columnwidth]{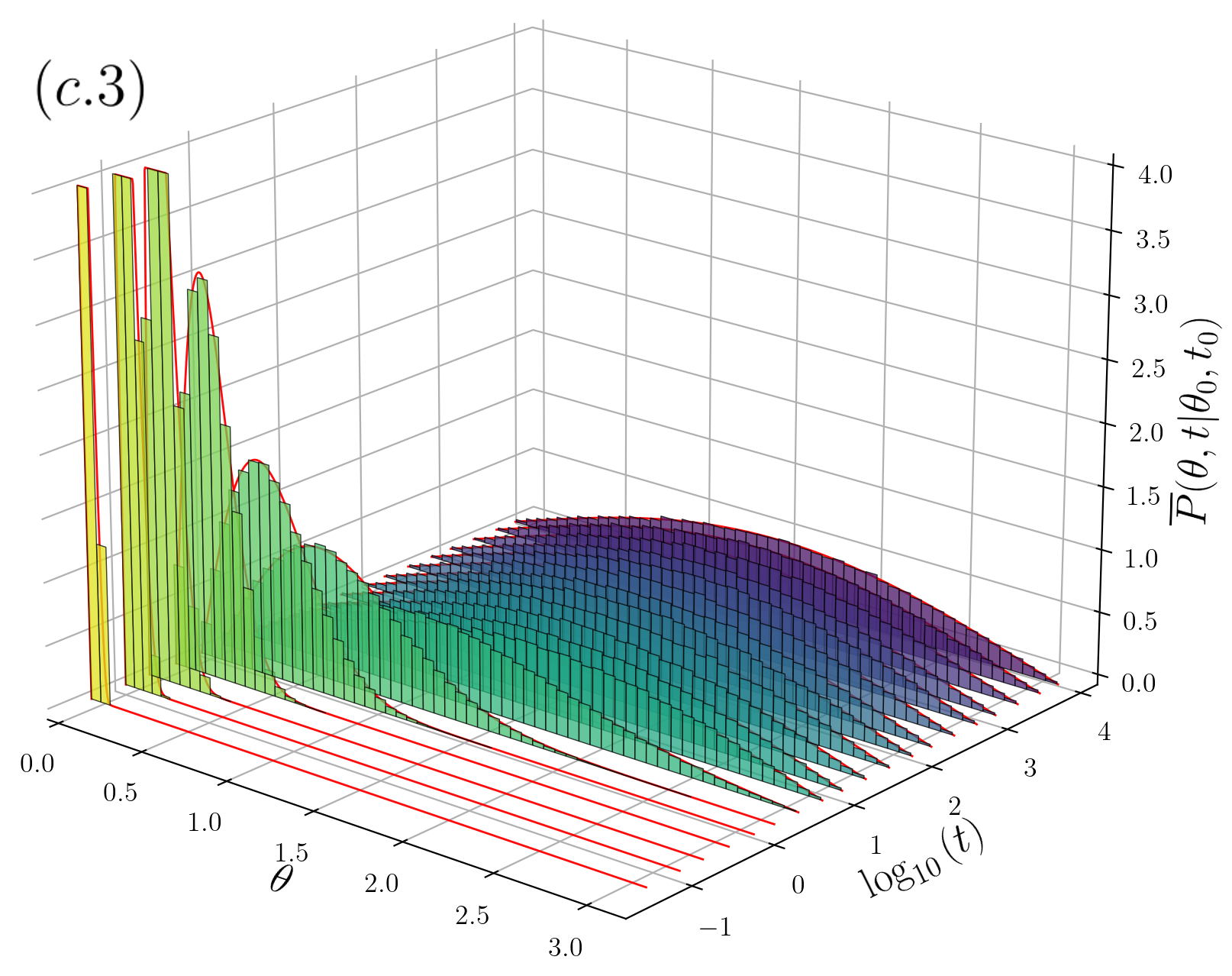}
\caption{The frequency distributions of the polar angle $\overline{P}(\theta,t\vert\theta_0,t_0)$ at different times ($\log t$) is shown for sBm. 
Numerals 1, 2 and 3 correspond to the values $H=0.1$, $H=0.5$ and $H=0.9$, respectively. In all cases we have set $\theta_0=0$ at $t_0=0$ and as for fBm, $H=0.5$ in [(c.2)], Brownian motion on $\mathbb{S}^2$ is recovered. A remarkable agreement between the distributions computed from the trajectories ensemble and the distributions obtained from the analytical solution  \eqref{PDF} (continuous-red lines) is shown.  $P_{st}(\theta)=\sin\theta/2$ is shown in red-dashed line (indistinguishable from the histograms computed in the long-time regime).}
\label{Theta-Time-Hist-sBm}
\end{figure}

Each ensemble of trajectories was generated with fixed initial position at the sphere's north pole $\boldsymbol{r}_{0}=(0,0,1)$, this choice allows us to assume an azimuthal symmetry in the ensemble average and therefore to focus on the frequency distribution of the polar angle $\theta$ only. From each ensemble, the frequency histograms of the polar angle $\overline{P}(\theta,t\vert\theta_0,t_0)${, with $\theta_0=0$ at $t_0=0$,} is computed at each time step {giving rise to} the time evolution shown in Figs.~\ref{Theta-Time-Hist} and \ref{Theta-Time-Hist-sBm}; in Fig. \ref{Theta-Time-Hist} we show the case for fractional Brownian motion, for which long-time correlations make manifest the role of {the} navigation strategy through the basis system chosen at each $\mathbb{T}_{\boldsymbol{r}}\mathbb{S}^2$. In Fig. \ref{Theta-Time-Hist-sBm} we show the corresponding histograms for the case of sBm.

For the Frenet-Serret navigation strategy $\{\hat{\boldsymbol{t}},\hat{\boldsymbol{n}}\}$ {(second column of Fig. \ref{Theta-Time-Hist})}, the system attains the stationary distribution $P_{eq}(\theta)$ {for each $H$}, {this} is shown in the evolution of the frequency histograms in {the} plots {of the second column of} Fig.~\ref{Theta-Time-Hist} for $H=0.1$ (b.1), 0.5 (b.2) and 0.9 (b.3) respectively ($P_{eq}$ is depicted by the dashed-red line). Notice that the relaxation times towards $P_{eq}(\theta)$ depend on $H$, being shorter the smaller $H$ is [compare the time scales considered {from (b.1) to (b.3) histograms in the second column of Fig.}~\ref{Theta-Time-Hist}]. In contrast, for the navigation strategy $\{\hat{\boldsymbol{\theta}},\hat{\boldsymbol{\phi}}\}$, the system attains a \emph{non-equilibrium stationary state} $P_\textrm{ness}(\theta)$ (except for the uncorrelated case $H=0.5$, for which the attained stationary state is independent of the navigation strategy and coincides with $P_{eq}(\theta)$ shown in the histograms of (a.2) of Fig.~\ref{Theta-Time-Hist}). For anti-correlated motion ($0<H<0.5$) the frequency histograms settle into a unimodal  distribution, highly peaked around the equator line ($\theta=\pi/2$) the smaller the value of $H$ [see histograms in (a.1) of Fig.~\ref{Theta-Time-Hist} for $H=0.1$], the distribution transforms continuously into $P_{eq}(\theta)$ as $H\rightarrow0.5$ (a.2). For correlated motion ($0.5<H<1$) {the stationary distribution} settles into a bimodal distribution with modes at the sphere poles [see histograms in (a.3) of Fig.~\ref{Theta-Time-Hist} for $H=0.9$], such stationary distribution is determined by the specific basis chosen only, and does not depend on the initial distribution. Thus, the poles can be interpreted as being \textit{repulsive} if $0<H<0.5$, and being \textit{attractive} if $0.5<H<1$. These non-equilibrium stationary distributions, when $H\neq 0.5$, are attained independently of the initial distribution.

Analogous effects to the ones observed with the basis induced by the spherical coordinates have been observed in confined fBm, specifically as result of the interplay of long-time correlations and hard-walls boundaries \cite{WadaPRE2018,GuggenbergerNJPhys2019,VojtaPRE2020}, where depleted regions for the probability distribution of the particle positions are observed near the boundaries for anticorrelated motion ($0<H<0.5$), while accreted regions near the boundaries are observed for $0.5<H<1$.

{In Fig.~\ref{Theta-Time-Hist-sBm} shows $\overline{P}(\theta,t\vert\theta_0,t_0)$ for sBm. $P_{eq}$ is attained for all $H$ and the cases $H=0.1$ (c.1), $0.5$ (c.2) and $0.9$ (c.3) are shown. Excellent agreement is observed except for the short-time regime of $H=0.1$ and 0.2, where there is some discrepancy due to numerical stability.}

\section{Comparative analysis of the cases considered.-}
\begin{figure*}
\includegraphics[width=0.33\textwidth,trim=0 0 20 20,clip=true]{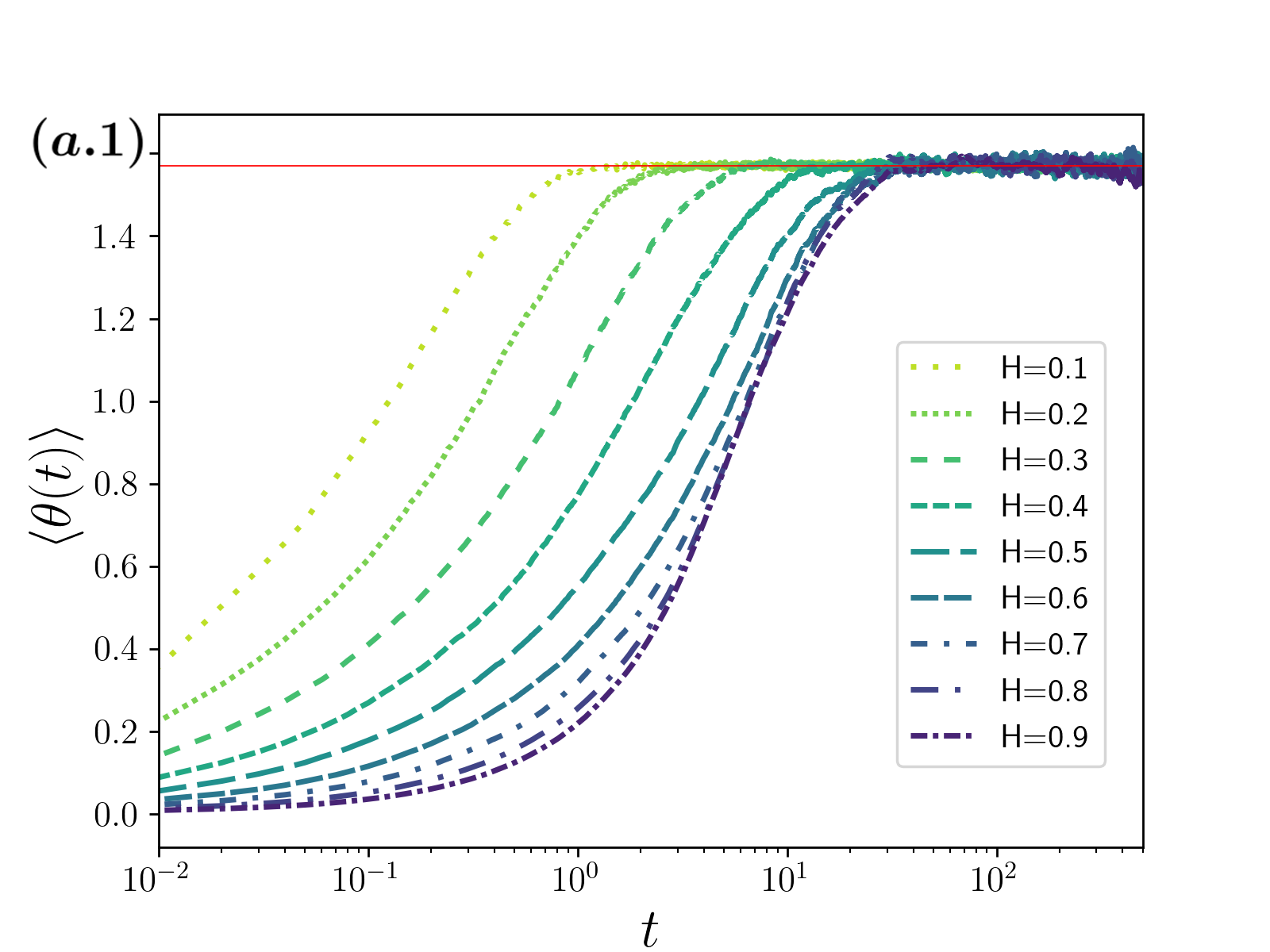}\includegraphics[width=0.33\textwidth,trim=0 0 20 20,clip=true]{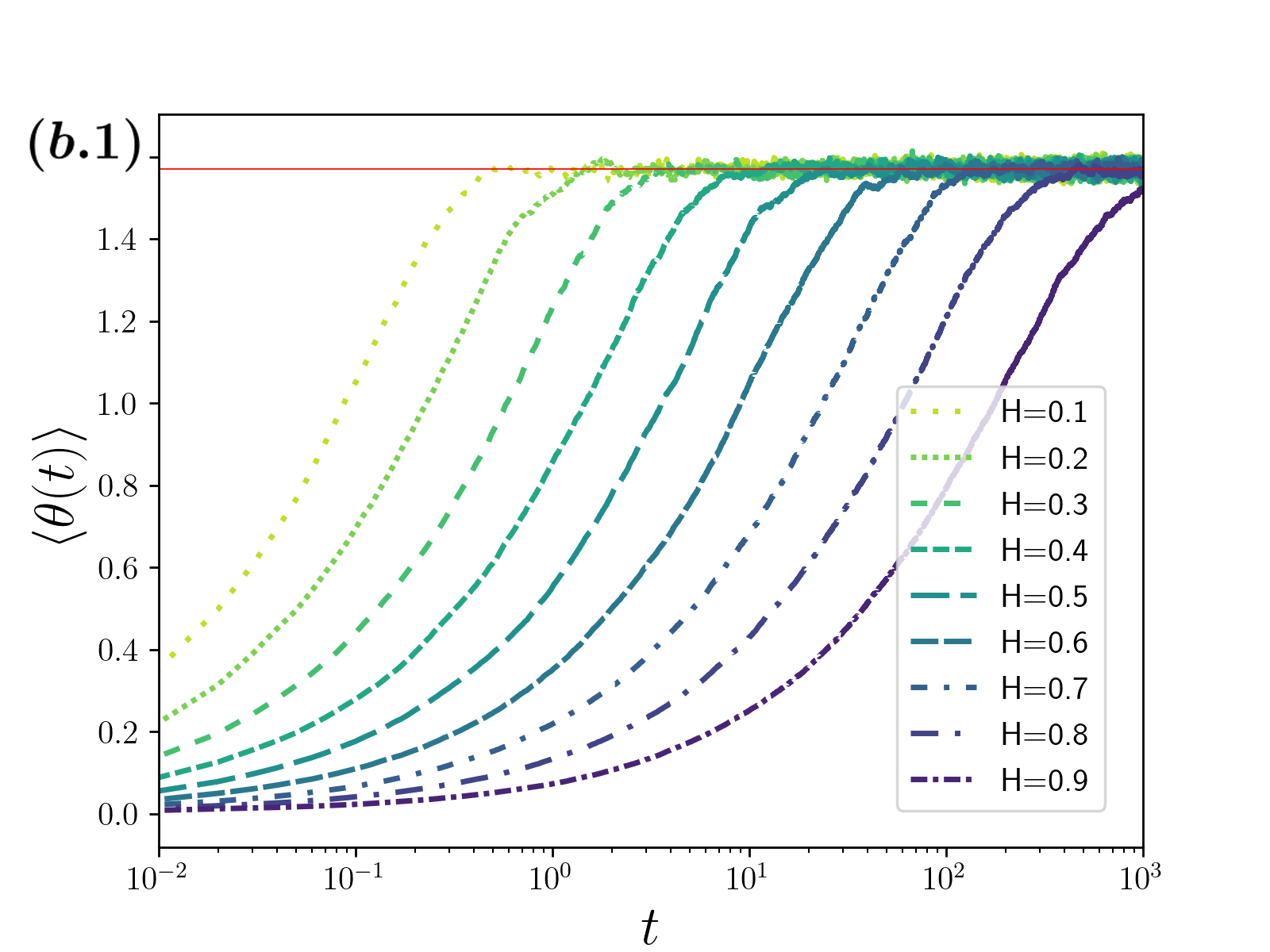}\includegraphics[width=0.33\textwidth,trim=0 0 20 20,clip=true]{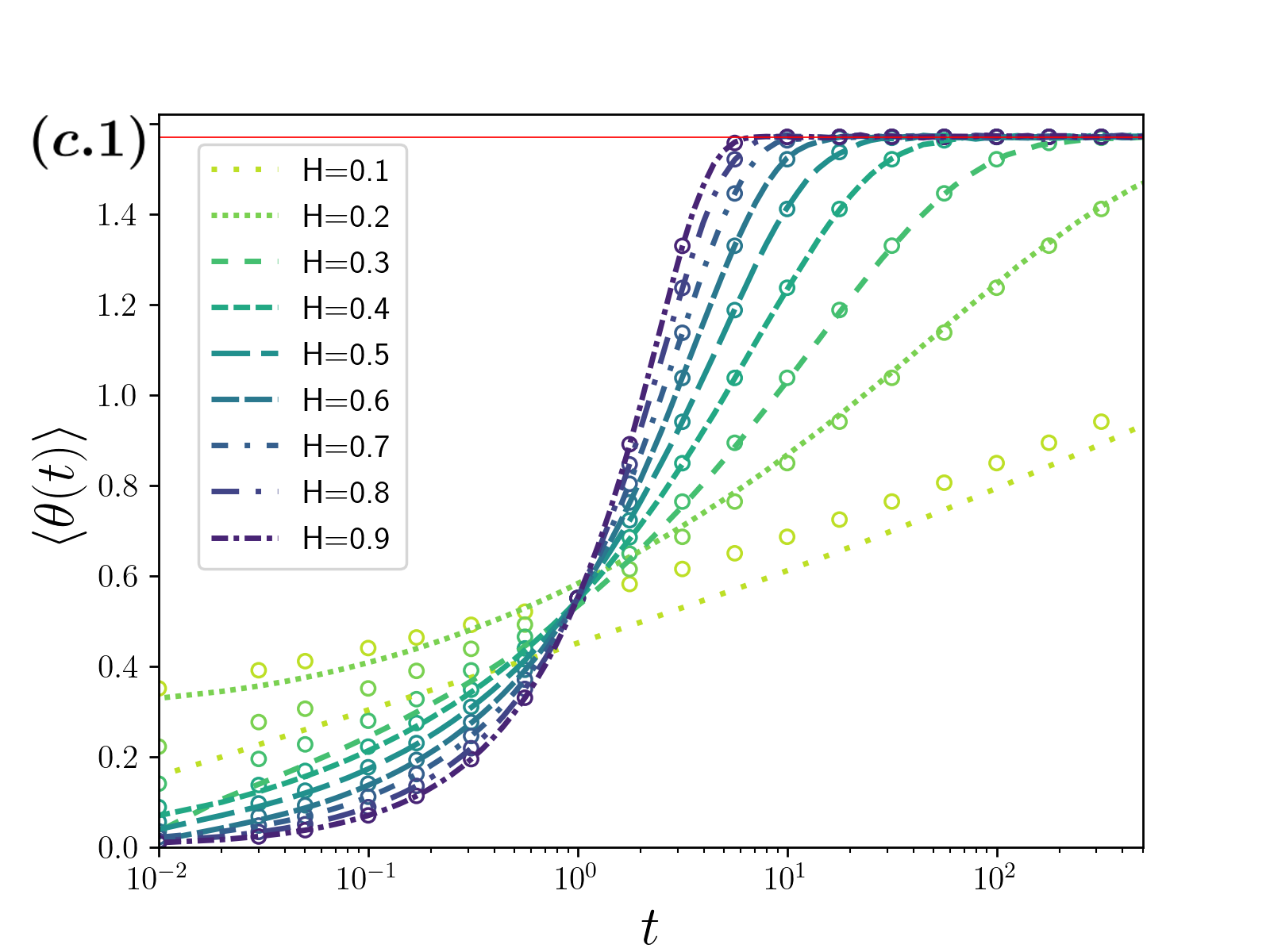}\\
\includegraphics[width=0.33\textwidth,trim=0 0 20 20,clip=true]{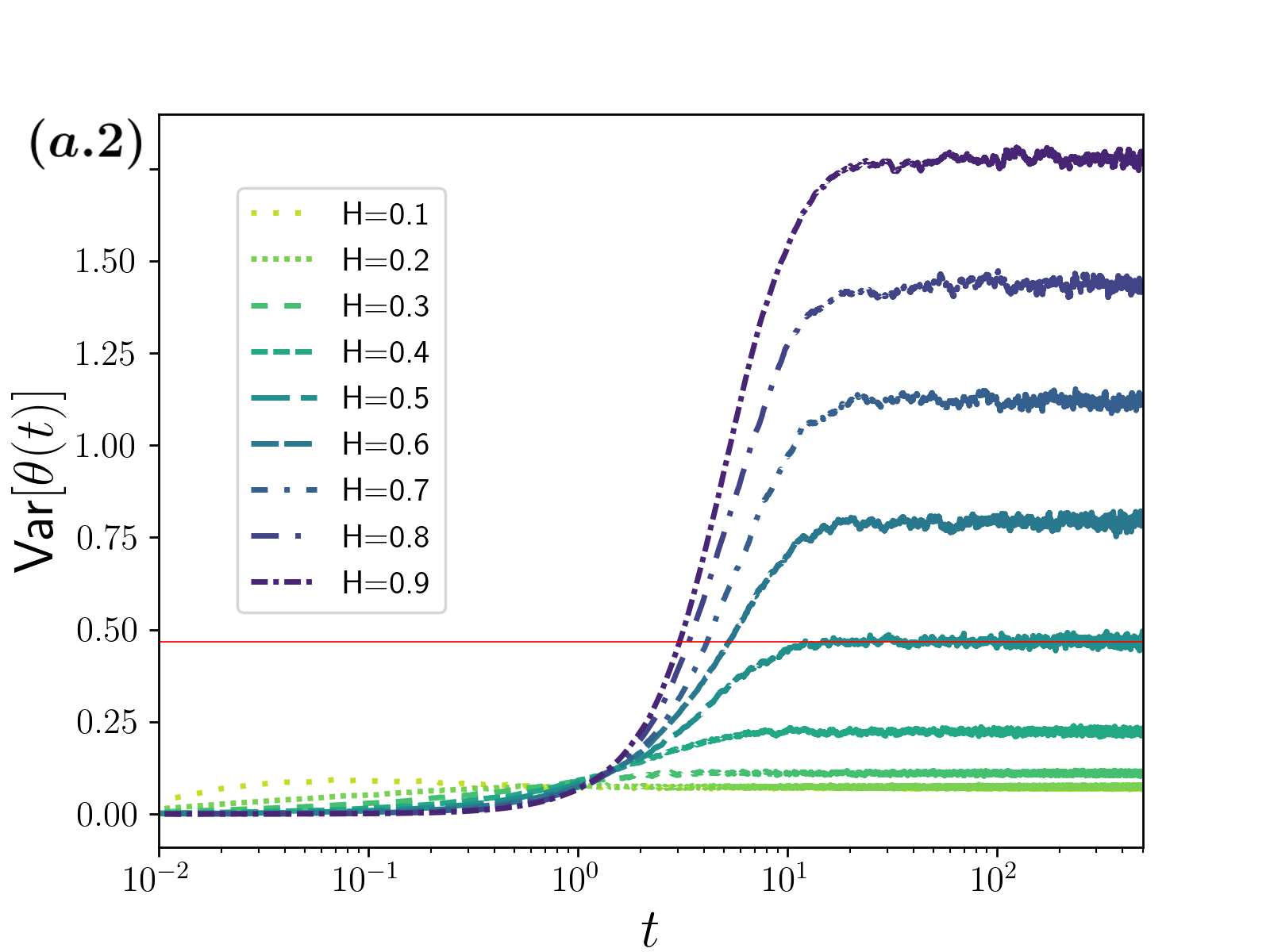}\includegraphics[width=0.33\textwidth,trim=0 0 20 20,clip=true]{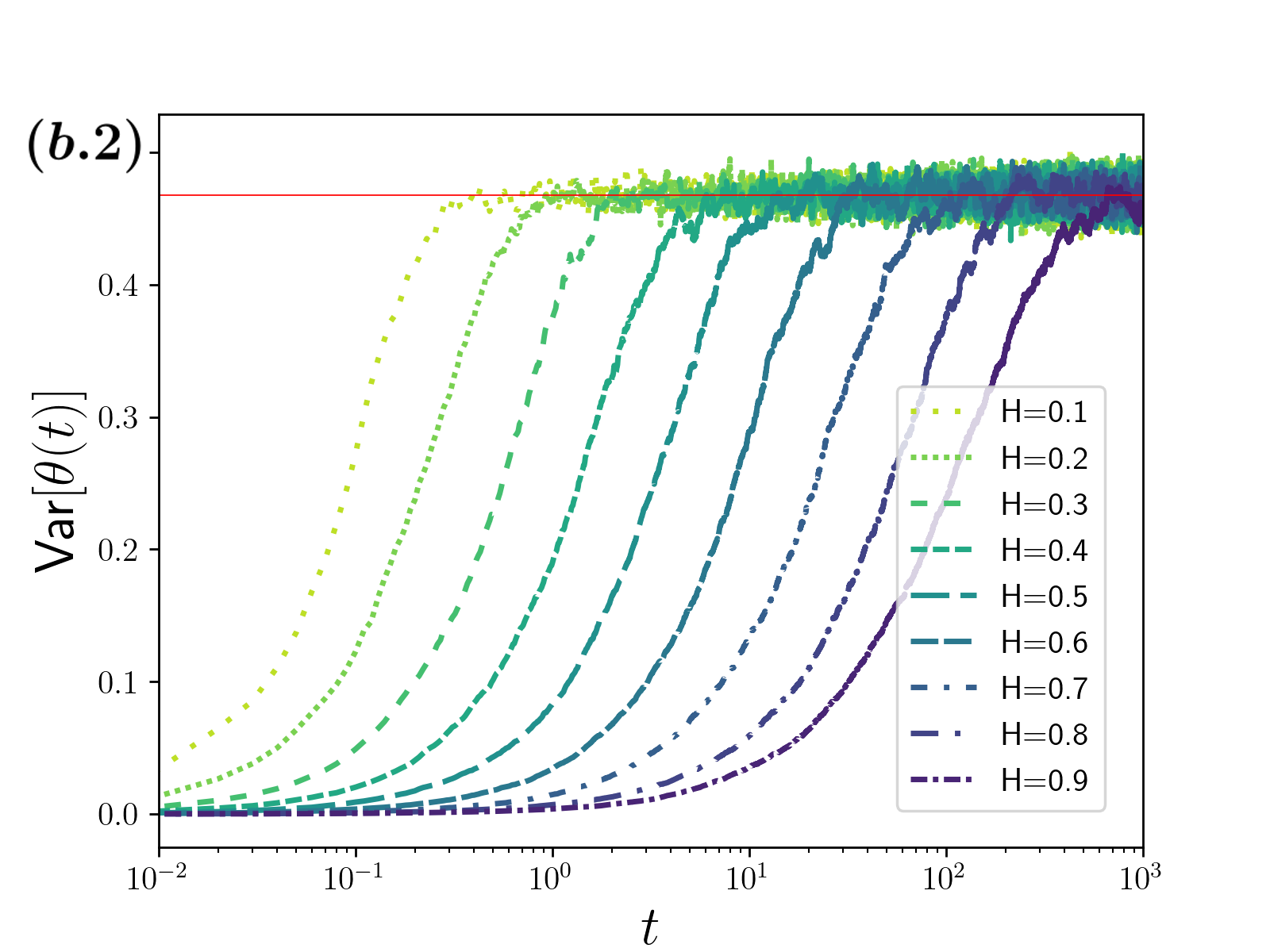}\includegraphics[width=0.33\textwidth,trim=0 0 20 20,clip=true]{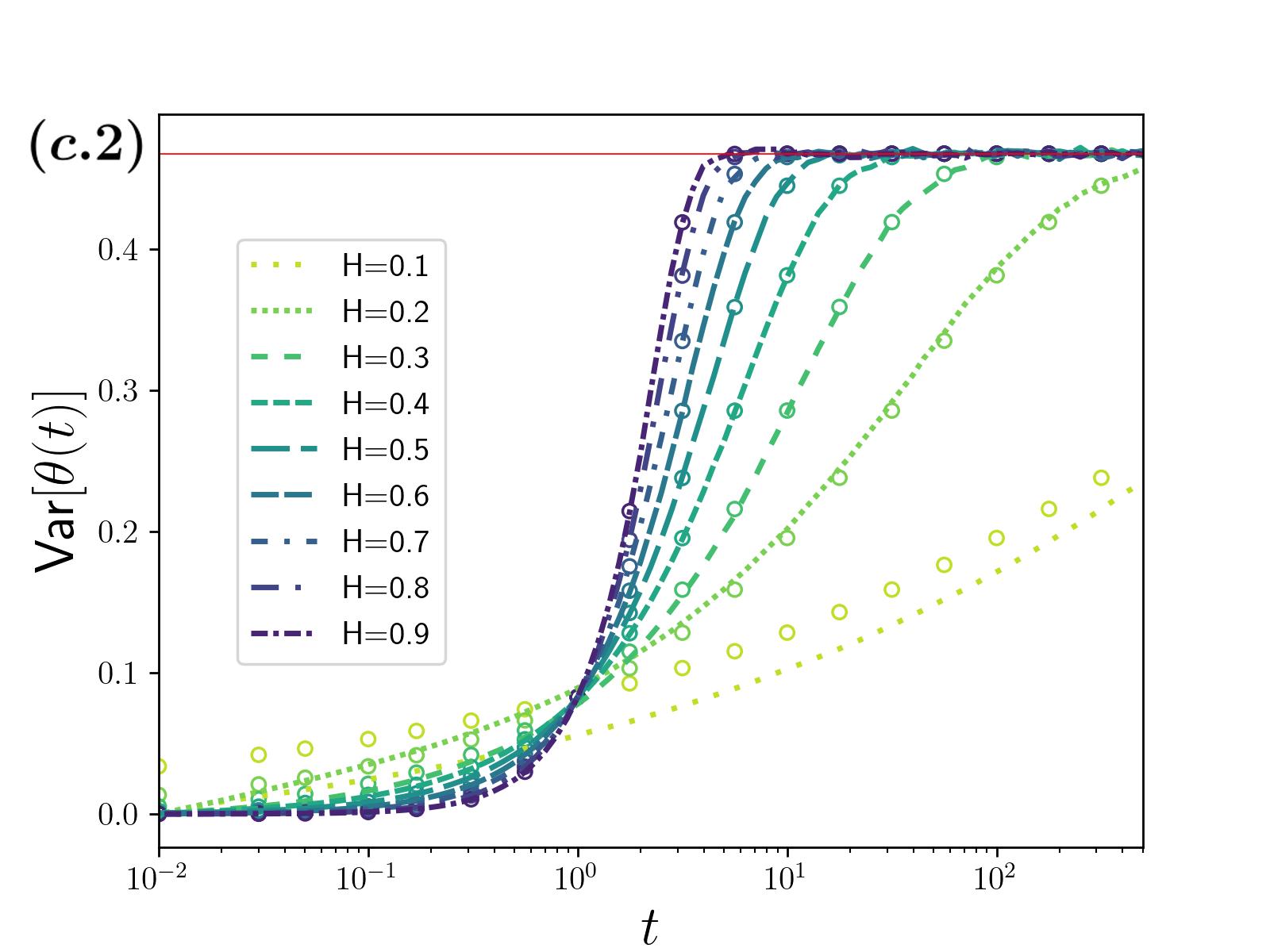}\\
\includegraphics[width=0.33\textwidth,trim=0 0 20 20,clip=true]{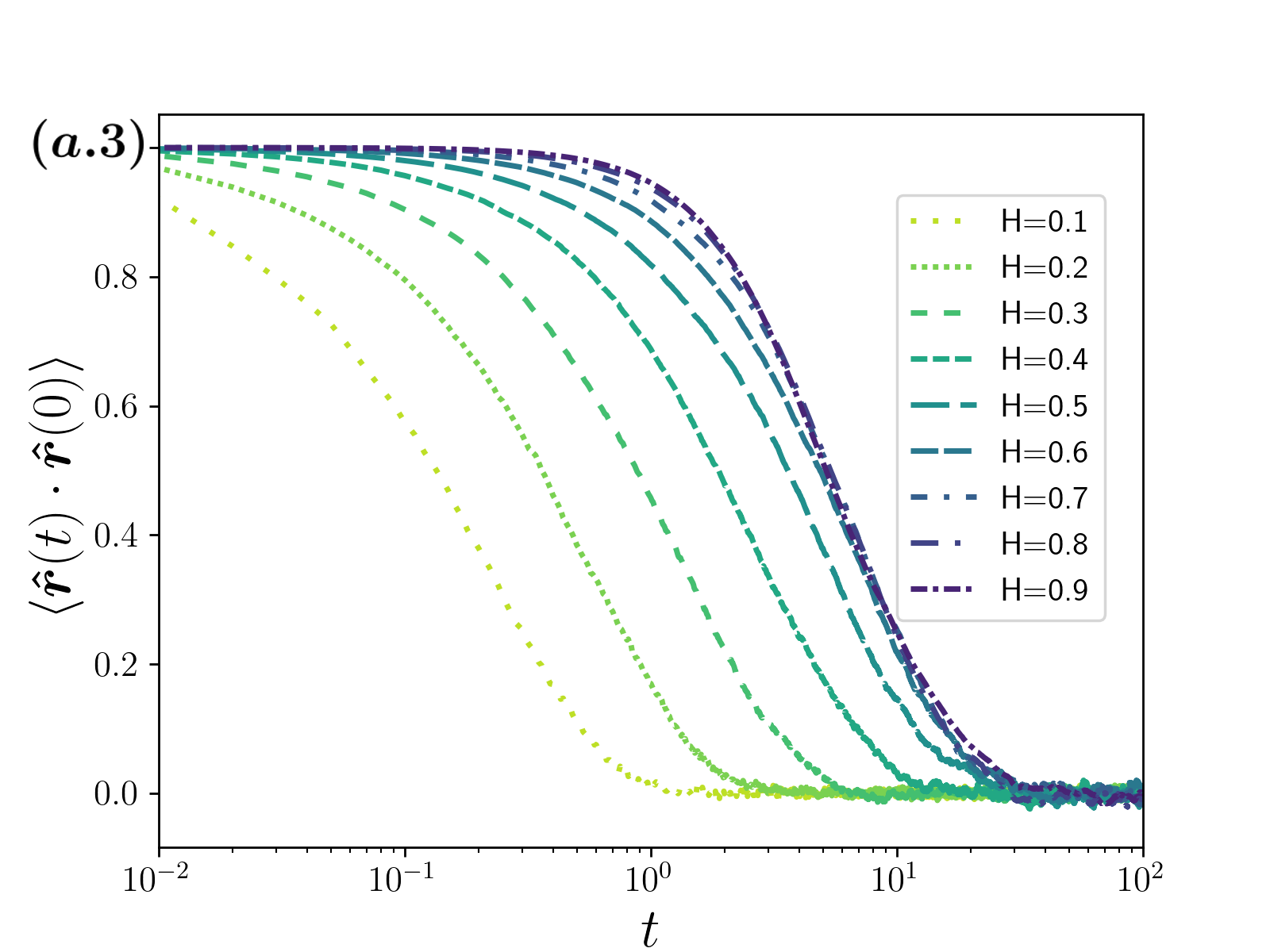}\includegraphics[width=0.33\textwidth,trim=0 0 20 20,clip=true]{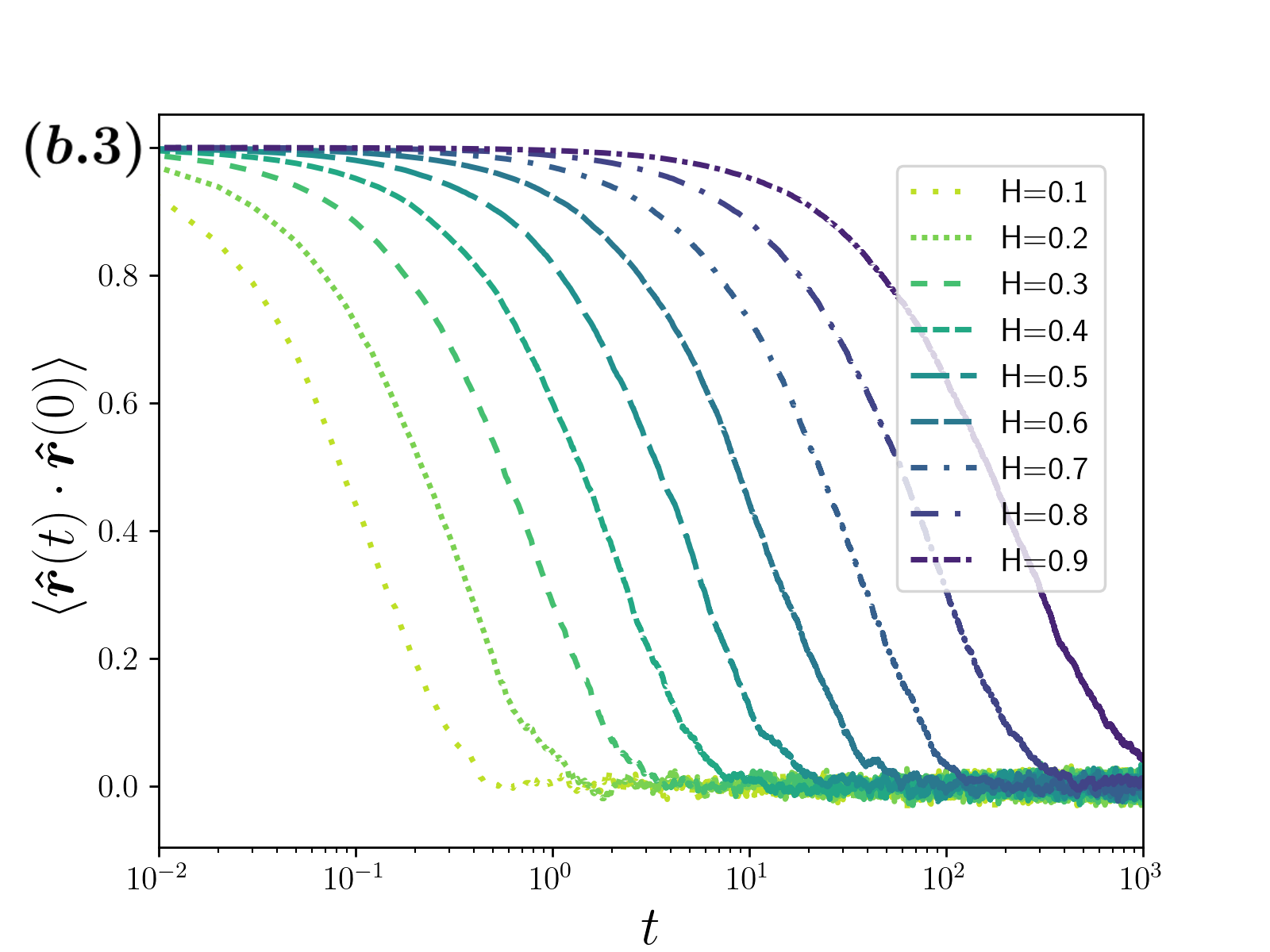}\includegraphics[width=0.33\textwidth,trim=0 0 20 20,clip=true]{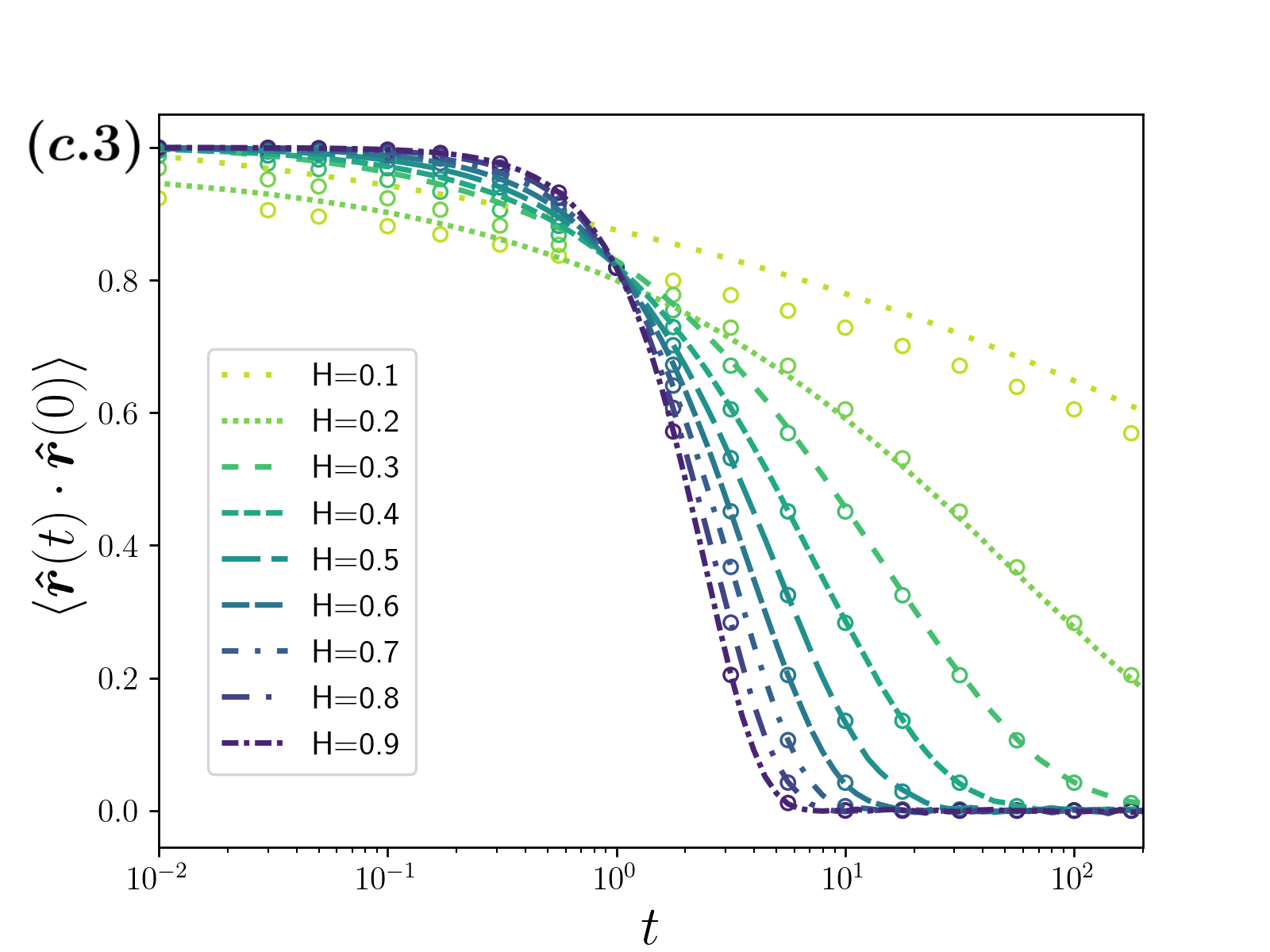}
\caption{Mean $\langle\theta(t)\rangle$, variance Var$[\theta(t)]$, and position autocorrelation function $\langle\hat{\boldsymbol{r}}(t)\cdot\hat{\boldsymbol{r}}(0)\rangle$ of the $\theta$ coordinate, for fBm with: navigation strategy given by the spherical coordinates basis $\{\hat{\boldsymbol{\theta}},\hat{\boldsymbol{\phi}}\}$ [(a) column], navigation strategy given by the Frenet-Serret basis $\{\hat{\boldsymbol{t}},\hat{\boldsymbol{n}}\}$ [(b) column], and for the scaled Brownian motion [(c) column]. For each case, the values $H=0.1$, 0.2, \ldots, 0.9 are considered. The amplitude of fluctuations for fBm, and the amplitude of the diffusion constant of sBm, both denoted with $D_H$ was set to 0.1 (arbitrary units). Continuous-thin red lines in (a.1). (b.1) and (c.1), indicate the values of $\langle\theta\rangle$ for the uniform distribution on the sphere $P_{st}(\theta)$; while they indicate the corresponding value for the variance in (a.2), (b.2) and (c.3) . In all cases, the lines refer to the results obtained from the ensemble of trajectories generated with our numerical method, while open dots in column (c) refer to the values computed from the analytical solutions \eqref{PDF}.}
\label{fig:Moments}
\end{figure*}

The qualitatively different behavior among the three cases considered is apparent from the analysis of the time dependence of three relevant statistical quantities of the distribution of the polar angle, namely: The mean $\langle\theta(t)\rangle$, the variance $\textrm{Var}[\theta(t)]=\langle\theta^{2}(t)\rangle-\langle\theta(t)\rangle^{2}$, and the position autocorrelation function $\langle\hat{\boldsymbol{r}}(t)\cdot\hat{\boldsymbol{r}}(0)\rangle$, {all of them} shown in Fig.~\ref{fig:Moments} and identified with the numerals 1, 2 and 3 correspondingly; the column (a) corresponds to the case of fBm with navigation strategy $\{\hat{\boldsymbol{\theta}},\hat{\boldsymbol{\phi}}\}$, the (b) one for fBm with $\{\hat{\boldsymbol{t}},\hat{\boldsymbol{n}}\}$ and the column (c) for sBm, which is independent of the navigation strategy. For each case considered we have $\langle\cdot\rangle\equiv\int_{0}^{\pi}d\theta\sin\theta\,(\cdot)P_\textrm{xBm}(\theta,t\vert 0,0)$, with xBm either fBm or sBm as appropriate.

Notorious differences caused by the navigation strategy in the time evolution of the quantities considered can be observed for fBm. On the one hand, in the case of the basis $\{\hat{\boldsymbol{\theta}},\hat{\boldsymbol{\phi}}\}$, $H=0.5$  separates two qualitatively distinct  behaviors. For $0<H<0.5$ the position autocorrelation decays faster than for the case $H=0.5$ the smaller $H$ is (see (a.3) in Fig.~\ref{fig:Moments}), as consequence, the mean approaches the value $\pi/2\approx1.5708$, also faster the smaller $H$ is (see (a.1) in Fig.~\ref{fig:Moments})]. For $0.5<H<1$ the correlations decay slower than for $H=0.5$, however, the time dependence in this case is similar. In each case the distributions are symmetric around the equator thus $\langle\theta\rangle$ saturates at $\pi/2$. The departure from the uniform distribution on $\mathbb{S}^2$ is evidenced by the time dependence of the variance [(a.2) in Fig.~\ref{fig:Moments}], which clearly indicates a smaller dispersion around the mean for $0<H<0.5$ (the distributions are more peaked around the mean with kurtosis $\kappa=3.0657$ for $H=0.1$, $\kappa=3.0054$ for $H=0.2$, $\kappa=2.8975$ for $H=0.3$ and $\kappa=2.6675$ for $H=0.4$, with respect to the uniform distribution, $H=0.5$, whose  kurtosis has the exact value $\kappa=2.19375$). Larger dispersion is observed for $0.5<H<1$ due to the bimodality of the distribution of $\theta$ (the variance of the distribution highly peaked at the poles, notice that the distribution $\frac{1}{2}[\delta(\theta)+\delta(\theta-\pi)]$ has variance $\pi^2/4\simeq2.4674$ which serves as an upper bound for the values presented in (a.2) of Fig.~\ref{fig:Moments}).

When the basis used is $\{\hat{\boldsymbol{t}},\hat{\boldsymbol{n}}\}$ the position autocorrelation decays with a relaxation time that increases monotonically with $0<H<1$ as is shown in (b.3) of Fig.~\ref{fig:Moments}. Similarly, the mean (b.1) and variance (b.2) saturate to the values of $P_{eq}(\theta)$ at a pace that depends monotonically on $H$, the smaller the $H$ the faster reaches the values corresponding to $P_{eq}(\theta)$.

It is clear from these results that for a basis pinned on the manifold induced by a set of local coordinates, as {for} the spherical coordinates $\{\hat{\boldsymbol{\theta}},\hat{\boldsymbol{\phi}}\}$ analyzed here, anti-correlated motion ($0<H<0.5$) makes the sphere poles to be seldom visited (the regions around the poles are diluted in the ensemble) in contrast, for correlated motion ($0.5<H<1$), the region around the sphere poles are frequently visited. It is only for $H=0.5$ (non-correlated motion) from which the uniform stationary distribution $P_{eq}(\theta)$ {is attained}, for which each location on the sphere should be visited as frequently as any other independently of the initial distribution. We thus conclude that long-correlated motion breaks the spherical symmetry {of the stationary state} for locally-coordinates induced basis $\{\hat{\boldsymbol{u}},\hat{\boldsymbol{v}}\}$. {In constrast,} fully spherical symmetry {of the stationary state is attained for all values of $H$} for {a} basis unpinned from the sphere local coordinates, {as shown here for} the Frenet-Serret system that serves as a body-reference system. 

Finally, we contrast the results obtained for sBm (column (c) in Fig.~\ref{fig:Moments}) with those presented for fBm. Since long-time correlations do not collude with navigation strategy in th{e sBm} case, $P_{eq}(\theta)$ is attained in the long-time regime for all $0<H<1$. The position autocorrelations decay faster the larger $H$ is, which is a consequence of the fact that the $P_{eq}(\theta)$ distribution of the polar angle relaxes faster with $H$ in contrast with fBm case. The time dependence of the mean [4~(c.1)], variance [4~(c.2)] and position autocorrelation function [4~(c.3)] are shown in the column (c) in Fig.~\ref{fig:Moments}, where calculations from the ensemble (lines) are compared with the analytical expression given by the the stretched exponential in Eq. \eqref{ACFfBm} (open dots). The discrepancy observed for the smaller values of $H$ is mainly due to the combined effects of the large fluctuations induced by the time dependence of the diffusion coefficient $D(t)$ in the short-time regime and size of our ensemble of trajectories. {We want to comment in passing that due to the $H$-invariance of sBm under the scaling transformation $\tau_{H}=D_H\, t^{2H}$, the curves presented in Fig. \ref{fig:Moments} will overlap on top of each other if plotted as function of $\tau_H$ instead of $t$.}

\paragraph*{Concluding remarks.-} We studied fractional Brownian motion as a model of long-time correlated motion of a particle that randomly swims on the surface of a sphere. We elucidated the deep connection {of this process} with {different} navigation strategies defined by {a} particular {basis for the tangent planes to $\mathbb{S}^2$} that specifies the updating rule {of the} algorithm {used} to analyze random motion on the sphere. In the case of uncorrelated motion ({for which} fBm with $H=0.5$ {is a particular case}), the dynamics is independent of the navigation strategy coinciding with standard Brownian motion on the sphere. {Furthermore,} we formulated {the updating rules in the algorithm used (Eqs.~\eqref{UpdatingRule}-\eqref{ScalingFactor})} to {analyze the} diffusion processes characterized by a time-dependent diffusion coefficient $D(t)$, and focused in scaled Brownian motion for which $D(t)=2HD_{H}t^{2H-1}$. 

We found that for the navigation strategy induced by the spherical coordinates, two qualitatively distinct trajectory patterns are observed depending {on whether} $0<H<0.5$ or $0.5<H<1$ which lead to non-equilibrium stationary states. The equilibrium stationary distribution is recovered  for the navigation strategy induced by the Frenet-Serret system which unpins the long-time correlated motion from a basis that depends on the local coordinates. These observations allow to conclude that long-time correlated motion collude with the class of navigation strategies that locally depend on the coordinates chosen, leading to ``nonequilibrium steady states'' breaking down equilibrium dynamics, this last one is restored for the class of coordinates-independent navigation strategies {(Brownian motion and sBm)}.

The similarity of the effects reported here for the coordinates-dependent navigation strategy in fBm (the sphere poles are seldom visited for $0<H<0.5$ while frequently visited for $0.5<H<1$), with those occurring in confined fBm in Euclidean space (depletion of the position distribution around reflecting walls for $0<H<0.5$ and accumulation for $0.5<H<1$) reported in Ref. \cite{VojtaPRE2020} are remarkable, and compels to conjecture that such similarity finds its origin in the coordinates-dependent navigation strategy chosen. It is plausible that if the body-reference system is employed in the dynamics in Euclidean space, instead of a fixed local space basis, it will restore a dynamics that leads to the uniform distribution in the long-time regime. Research in this direction has been started and the corresponding results will be published elsewhere.

{To conclude} we can classify the navigation strategies into two main classes: Those that are pinned to the manifold by a particular choice of coordinates {basis for each tangent plane $\mathbb{T}_{\boldsymbol{r}_n}\mathbb{S}$}; and those that are fixed at the body-reference system. Both situations are plausible to occur in real systems. For instance, {the lateral movement of organic molecules on the cell membrane,} {are} subject to {a variety of effects induced by the} surface {(many lipids distributed on the membrane serve as obstacles)} on which the {molecules} move {(this would lead to a navigation strategy pinned to the manifold).} 
This is the case for the lateral movement of signaling phosphatidylinositol lipids
(PIP2, PIP3, PIP \cite{McLaughlin.2002,Beard.2008,Hancock.2010}) on the cell membrane. Diffusion of these molecules is affected by the interaction with the cell membrane lipids and other of its components \cite{Yeagle.2005,Brough.2005, FujiwaraJCellBio2002,BlindPNAS2014,BucklesBioPhysJ2017}, thus any mobility correlations arise from the distribution of obstacles in the cell membrane surface, and therefore can be modeled by use of a reference frame fixed in the cell membrane. These same effects make water molecules to subdiffuse on the cell membrane \cite{YamamotoSciReps2014}.
With the recent advances designing cell membranes coated particles \cite{SpanjersAdvBiosystems2020}, the diffusion of proteins on such surfaces can be modified either by the shape of the coated particle or by the properties of the coating membrane. {In the same trend and with the aim to overcome the obstacles faced by traditional synthetic micromotors, active (or self-propelling) micromotors have been cell membrane-functionalized to enhance mobility due to motility induced persistence \cite{Esteban-FernandezAccChemRes2018,WuSciAdv2020, ZhangAdvMaterials2022} these are able to navigate tissue,} subject to an ``internal'' navigation scheme {(navigation strategy that depends on the body-reference)} that  arise from {the} self-propulsion mechanisms.{The effects of fBm on the self-propulsion dynamics of active particles moving in two-dimensional space have been analyzed in Ref. \cite{Gomez-SolanoJStatMech2020}.}  

{Finally, single-tracking methods have provided evidence that motion in crowded environments, particularly molecular motion in the cell membrane (transmembrane motion), is not as simple as Brownian motion \cite{RitchieBioPhysJ2005,HoflingRepProgPhys2013,MetzlerBiochBiophysActa2016}. This has led to the formulation of a variety of diffusion models to capture the anomalous diffusion behavior of these elements. Those based on fractional Brownian motion have shed light in the understanding of anomalous diffusion in numerous biophysical systems \cite{BenelliNJPhys2021,ErnstSoftMatter2012,SadoonPhysRevE2018,SabriPhysRevLett2020,SpecknerEntropy2021}.}

In this paper we have highlighted the non-trivial effects of the interplay between the reference frame, that describe the motion of particles moving on the surface of the sphere, and the long correlations of the motion. The analysis presented has potential applications to biophysical systems, particularly in the analysis of the diffusion of molecules in biological membranes, where the motion, besides being stochastic, is in many cases long-time correlated.


\acknowledgments
This work was supported by UNAM-PAPIIT IN112623. 

\appendix

\end{document}